\documentclass[aps,twocolumn,pra,superscriptaddress]{revtex4-2}

\usepackage{multirow}
\usepackage{booktabs}

\usepackage[T1]{fontenc}
\usepackage{dsfont}
\usepackage{amsfonts}
\usepackage{amsmath}
\usepackage{bm}
\usepackage{amssymb}
\usepackage{graphicx}
\usepackage{braket}

\usepackage[colorlinks=true,urlcolor=blue,citecolor=blue,linkcolor=blue]{hyperref}
\DeclareGraphicsExtensions{.eps,.ps,.pdf}
\newcommand{\diff}{\mathop{}\!\mathrm{d}}
\newcommand{\al}{\alpha}

\newcommand{\la}{\lambda}
\newcommand{\ka}{\kappa}

\newcommand{\si}{\sigma}

\newcommand{\La}{\Lambda}

\newcommand{\TRS}{\mathcal{T} }

\newcommand{\BM}{\begin{displaymath}}

\newcommand{\EM}{\end{displaymath}}

\newcommand{\ie}{\hbox{\em i.e.{}}}

\def \be  {\begin{equation}}
\def \eeq {\end{equation}}
\def \baa {\begin{eqnarray*}}
\def \eaa {\end{eqnarray*}}
\def \bb  {\begin{thebibliography}}
\def \eb  {\end{thebibliography}}

\def \lab #1 {\label{#1}}

\def \Tr {\text{Tr}}

\newcommand{\Rr}{\mathsf{R}}

\newcommand{\co}{\mathcal{C}}

\newcommand{\da}{\dagger}

\newcommand{\con}[1]{\mathcal{C}_{#1} }

\newcommand{\gf}{\gamma}

\makeatletter
\g@addto@macro\bfseries{\boldmath}
\makeatother

\usepackage[dvipsnames]{xcolor}

\usepackage[normalem]{ulem}
\usepackage{xcolor}
\usepackage{soul}
\setstcolor{red}
\usepackage{enumitem}   
 
 \newcommand\redout{\bgroup\markoverwith
{\textcolor{red}{\rule[.5ex]{2pt}{2.4pt}}}\ULon}
 \newcommand\blueout{\bgroup\markoverwith
{\textcolor{blue}{\rule[.5ex]{2pt}{2.4pt}}}\ULon}

\begin{document}
\widetext

\title{Phase characterization of spinor Bose-Einstein condensates: a Majorana stellar representation approach}

\author{Eduardo Serrano-Ens\'astiga}
\email[Corresponding author: ]{ed.ensastiga@uliege.be}
\affiliation{Departamento de F\'isica, Centro de Nanociencias y Nanotecnolog\'ia, Universidad Nacional Aut\'onoma de M\'exico\\
Apartado Postal 14, 22800, Ensenada, Baja California, M\'exico}
\affiliation{Institut de Physique Nucléaire, Atomique et de Spectroscopie, CESAM, University of Liège
\\
B-4000 Liège, Belgium}

\author{Francisco Mireles}
\email{fmireles@ens.cnyn.unam.mx}
\affiliation{Departamento de F\'isica, Centro de Nanociencias y Nanotecnolog\'ia, Universidad Nacional Aut\'onoma de M\'exico\\
Apartado Postal 14, 22800, Ensenada, Baja California, M\'exico}

\date{\today}
\begin{abstract}
We study the variational perturbations for the mean-field solution of an interacting spinor system with underlying rotational symmetries. An approach based upon the Majorana stellar representation for mixed states and group theory is introduced to this end. The method reduces significantly the unknown degrees of freedom of the perturbation, allowing us a simplified and direct exploration on emergent physical phenomena. We apply it to characterize the phases of a spin-1 Bose-Einstein condensate and to study the  behavior of these phases with entropy. The spin-2 phase diagram was also investigated within the Hartree-Fock approximation, where a non-linear deviation of the cyclic-nematic phase boundary with temperature is predicted. 
\end{abstract}

\maketitle
\section{Introduction}
\label{sec.Int}
Many-body quantum systems of interacting spins may exhibit novel phases and fascinating physical phenomena. In particular, in the field of ultracold atoms, phases  occurring in spinor Bose-Einstein condensates (BEC) can be realized and manipulated under highly controllable setups 
\cite{lewenstein2012ultracold,Kaw.Ued:12,PhysRevLett.119.050404,schafer2020tools,PRL.124.043001}. Most notably, the spin domain behavior of spinor BECs can differ drastically over different atomic species \cite{Kaw.Ued:12}. 
For instance, it has been corroborated experimentally~\cite{stenger1998spin,stamper.Andrews.etal:1998,Jacob:12,Chang.Hamley.etal:2004} that condensates of $^{23}$Na and $^{78}$Rb in an optical trap exhibits different ground spinor phases: the polar (P) and the ferromagnetic (FM) phases, respectively \cite{pethick2008bose,lewenstein2012ultracold}. 
Recent experimental advances allow us to scrutinize the spin phases of BEC of several spin values, from 1 to 8, even in the presence of external fields and spin-orbit interactions \cite{stamper.Andrews.etal:1998,jimenez2019spontaneous,Schmal.Erhard.etal:2004,PhysRevLett.104.063001,PhysRevLett.96.143005,schreck2021laser,PhysRevA.77.061601}. Theoretically, the study of the spin-phase diagram in spinor BECs has been done using mean-field (MF) theories, first for spin $f=1$ \cite{PhysRevLett.81.742,ohmi1998bose}, and subsequently for higher spins  \cite{Kaw.Ued:12,ciobanu2000phase,Bar.Tur.Dem:06,diener2006cr,PhysRevA.84.053616}. MF theory  
commonly reduces a many-body interacting problem into an effective one-body problem. This is achieved through the replacement of all interactions with an average over the many-body quantum states. Consequently, MF theory in BECs assumes that all the atoms in the condensate share the same single quantum state, such state characterized by a spinor order-parameter $\bm{\Phi}$ obeying the spinor version of the well-known Gross-Pitaevskii (GP) equations \cite{Kaw.Ued:12,pethick2008bose,lewenstein2012ultracold}.

Although the MF theory predicts qualitatively well the spin phases of a BEC, it fails to offer a satisfactory description of a wide range of physical effects as, its behavior at finite temperatures, quantum fluctuations, or non-local perturbations. The studies of the spinor BEC covering these aspects become essential to scrutinize other nontrivial physical phenomena such as deviations in the spin phase boundaries, metastable phases, tunneling effects, quench dynamics, or (dynamic/static) quantum phase transitions, among other phenomenology~\cite{Kawa.Phuc.Blakie:2012,Jacob:12,PRA.90.023610,PRA.95.053638,roy2022collective,Ser.Mir:21,PhysRevLett.78.3594,PhysRevA.61.063613,PhysRevA.74.033612,PhysRevA.78.023632,PhysRevA.73.013629,PhysRevA.88.043629,jimenez2019spontaneous,PRL.110.165301,NJP.095003,PRA.98.063618,PhysRevA.99.023606,PRA.100.013622,PRA.100.013622,PRL.124.043001}.

Some of the well-known beyond mean-field theories are the variational approaches, which have already proven to be well suitable near the MF phases in BECs~\cite{blaizot1986quantum,griffin2009bose,Kawa.Phuc.Blakie:2012,Pro2008,Gar.Mic.Cir.Lew:97,PhysRevA.102.043302,kanjilal2022variational}. The condensate gas, represented by a mixed ensemble of particles, is described by a density matrix $\rho$. It is assumed to be comprised by two contributions such that $\rho = \rho^c + \rho^{nc}$, where $\rho^c= \bm{\Phi}^{\da} \bm{\Phi}$ is the atom fraction that remains in the same MF solution $\bm{\Phi}$, while $\rho^{nc}$ is the ensemble of noncondensate atoms described by other quantum states to be determined~\cite{griffin2009bose,Kawa.Phuc.Blakie:2012,Ser.Mir:21}. 
Operationally, the study of the condensate and its perturbation entails the solution of the GP equations solved in a self-consistent manner with a set of equations that govern the noncondensate fraction~\cite{blaizot1986quantum,griffin2009bose}. Such approach is   computationally  demanding and not free from numerical issues. A way to circumvent this is to make use of variational methods exploiting the so called, \emph{self-consistent symmetries}~\cite{blaizot1986quantum}, \ie, approaches where the noncondensate fraction $\rho^{nc}$ inherits the common symmetries between the Hamiltonian of the whole cold gas and the order parameter $\bm{\Phi}$ of its condensed fraction. 

In fact, it is known that, as a consequence of the Michel's theorem~\cite{RevModPhys.52.617}, the common symmetries of the Hamiltonian of the
spinor sector of the BEC and $\bm{\Phi}$ usually conform a nontrivial point group, composed of a set of rotational and reflection symmetries. This result has been used to characterize MF solutions of spinor BEC~\cite{Bar.Tur.Dem:06,Mak.Suo:07,PhysRevA.75.023625,Kaw.Ued:12,PhysRevA.84.053616}. The point group symmetries associated to spin phases are also of great interest due to its connection to the appearance of Abelian and non-Abelian vortices~ \cite{PhysRevLett.89.190403,kasamatsu2005vortices,Koba.Kawa.Nitta.Ueda:2009}. Notably, the inherited symmetries of $\rho^{nc}$ has been exploited before to study the metastable phases of spinor BEC of spin-1 at finite temperatures~\cite{Ser.Mir:21}.

In this Letter, we present a thoroughly systematic method based on the Majorana representation of spin mixed states~\cite{Maj.Rep,Ser.Bra:20} and the use of {\it self-consistent symmetries}, that allows us   the full determination of the non-condensed fraction $\rho^{nc}$ of a spinor BEC having a particular point group. Our approach greatly reduces the number of degrees of freedom of the variational perturbation, offering a simplified methodology for the exploration of the nature of phase diagrams in BECs. We make use of the  approach to characterize the $\rho^{nc}$ of the spin phases of BECs of spin $f=1$ and $2$. In addition, as an application of these characterizations, we develop a simple  analytical model to study the spin-1 phases of a BEC under the presence of perturbations encoded by an increment in its entropy. The simple model reproduces well the known results for the phase diagrams reported earlier in the literature for BECs at finite temperatures. Furthermore, we study the phase diagram of the spin-2 BEC at finite temperatures using the Hartree-Fock approximation~\cite{blaizot1986quantum,griffin2009bose,Kawa.Phuc.Blakie:2012,Pro2008}, leading us to predict the appearance of a non-linear deviation with temperature of the cyclic-nematic phase boundary. 
\section{Methodology}
\label{Sec.Met}
We start by considering a BEC with spin $f$ confined in an optical trap. The atomic gas is assumed to be weakly interacting and sufficiently diluted such that only two-body collisions are predominant and that the $s$-wave approximation is still valid. Moreover, we will work in the regime where neither spin-orbit coupling nor the dipolar interactions are of significance. We also consider that the Hamiltonian model of the atomic gas is factorizable into its spinorial and spatial sectors, and that the BEC is free of any topological spin disorder.

Under such conditions, the spinor sector of the full Hamiltonian can be written in the second-quantization formalism in terms of the spinor field operators $\hat{\psi}_m$, and the numerical tensors associated to the two-body collisions, $\mathcal{M}^{(\gf)}$, \begin{equation}
\label{Full.Ham}
\hat{H} 
= \sum_{\gf=0}^{f} \sum_{i,j,k,l} \frac{c_{\gf}}{2} \mathcal{M}^{(\gf)}_{ijkl} \hat{\psi}_i^{\da} \hat{\psi}_j^{\da} \hat{\psi}_k \hat{\psi}_l  \, ,
\end{equation}
where the indices $i,\, j ,\, k ,\, l$ run over the magnetic quantum numbers $m= f,f-1, \dots , \, -f$ and $c_{\gf}$ are the coupling factors associated to the  $s$-wave scattering lengths of a given interaction channel, with  $\gf=0,\, 1, \dots f$ \cite{Kaw.Ued:12,lewenstein2012ultracold}. 
For example, the interactions for a spinorial BEC with $f=2$ has only three different   tensor elements, and are given by~\cite{Kaw.Ued:12} 
\begin{align}
\mathcal{M}^{(0)}_{ijkl} = & \delta_{il} \delta_{jk} \, , \quad 
\mathcal{M}^{(1)}_{ijkl} = (F_{\nu})_{il} (F_{\nu})_{jk}
\, ,
\nonumber
\\
& \mathcal{M}^{(2)}_{ijkl} = \frac{(-1)^{i+k}}{5}  \delta_{i,-j} \delta_{k,-l} 
\, ,
\end{align}
where $\delta_{ij}$ is the Kronecker delta, and $F_{\nu}$ are the components of the angular momentum matrices along the $\nu=x,y$ or $z$ direction, which here are scaled by $\hbar$ making them dimensionless. For spin-1 condensates, only the first two interactions $c_0$ and $c_1$ appear in the Hamiltonian. The  $c_0$-interaction is spin-independent since it is equivalent to the square of the number operator. The rest of the interactions are all spin-dependent. The spinor-quantum field operator associated to the spinor condensate is denoted by  $\hat{\bm{\Psi}}= (\hat{\psi}_f \, , \hat{\psi}_{f-1} \, , \dots \, , \hat{\psi}_{-f})^{\text T}$, 
where T denotes the transpose. Note that the Hamiltonian \eqref{Full.Ham} has a symmetry point group $SO(3) \times \mathds{Z}_2 $ constituted by the group of rotations $SO(3)$ and the inversion with respect to the origin.

Mean-field (MF) approximation assumes that  $\langle \hat{\bm{\Psi}} \rangle = \bm{\Phi}$, where $\bm{\Phi}= (\phi_f ,  \phi_{f-1} , \dots , \phi_{-f})^{\text{T}}$, is the spinor order-parameter obeying $\bm{\Phi}^{\da} \bm{\Phi} =N$, being $N$ the total number of atoms in the condensate gas~\cite{Kaw.Ued:12,lewenstein2012ultracold}. The spin phase of the BEC is thus the order parameter $\bm{\Phi}$ that minimizes the energy functional  $ E[\bm{\Phi}]= \langle \hat{H} \rangle$. The rotational symmetries of $\bm{\Phi}$ can be found through the Majorana representation for pure states \cite{Maj.Rep}, which associate each spin-$f$ state $\bm{\Phi}$ with $2f$ points on the sphere, usually called the (Majorana) constellation of $\bm{\Phi}$ and denoted by $\con{\bm{\Phi}}$. The representation is defined via a polynomial that involves the coefficients of $\bm{\Phi}$,
\begin{equation}
\label{first.pol}
p_{\bm{\Phi}}(z) = \sum_{m=-f}^f (-1)^{f-m}   \sqrt{\binom{2f}{f-m}} \phi_m z^{f+m}  \, .
\end{equation}
The degree of the polynomial $p_{\bm{\Phi}}(z)$ is at most $2f$. By rule, its set of roots $\{\zeta_k \}$ is always increased to $2f$ by adding the sufficient number of roots at infinity \cite{Maj.Rep,Chr.Guz.Ser:18}. $\co_{\bm{\Phi}}$  thus represents a set of $2f$ points on the sphere $S^2$, called \emph{stars}, and they are obtained through the stereographic projection from the south pole. Hence, each root $\zeta_k= \tan(\theta_k/2) e^{i \varphi_k}$ is associated to a point on $S^2$ with spherical angles $(\theta_k \, , \varphi_k)$. When $\bm{\Phi}$ is transformed by the unitary representation $D(\Rr )$ of a rotation $\Rr \in SO(3)$ in its Hilbert space, the constellation $\co_{\bm{\Phi}}$ rotates by $\Rr$ on the physical space $ \mathds{R}^3$, where $D^{(\si)}_{ \mu' \mu} (\Rr) \equiv \braket{\si , \, \mu' |     e^{-i \al F_z} e^{-i \beta F_y} e^{-i \gamma F_z} | \si , \, \mu}$ is the Wigner D-matrix of a rotation $\Rr$ with Euler angles $(\al, \, \beta , \, \gamma)$  \cite{Var.Mos.Khe:88}. Therefore, the point group of the quantum state $\bm{\Phi}$ is equal to the point group of the geometrical object associated to $\con{\bm{\Phi}}$. This representation has been used successfully to classify the spin ground phases of BEC in the ideal case of zero temperature~\cite{Bar.Tur.Dem:06,Mak.Suo:07,PhysRevA.84.053616}.

\begin{figure}[t]
\large
\begin{tabular}{|c|c|}
\hline
 $\mathcal{C}_{\bm{\Phi}} $ & $\mathcal{C}_{\rho^{nc}}$
\\  \hline
\scalebox{0.25}{\includegraphics{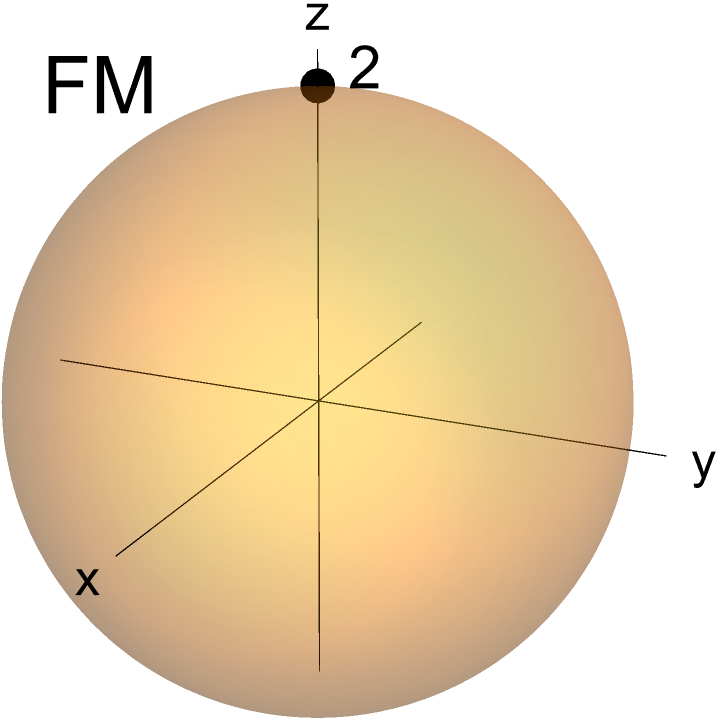}} & \scalebox{0.22}{\includegraphics{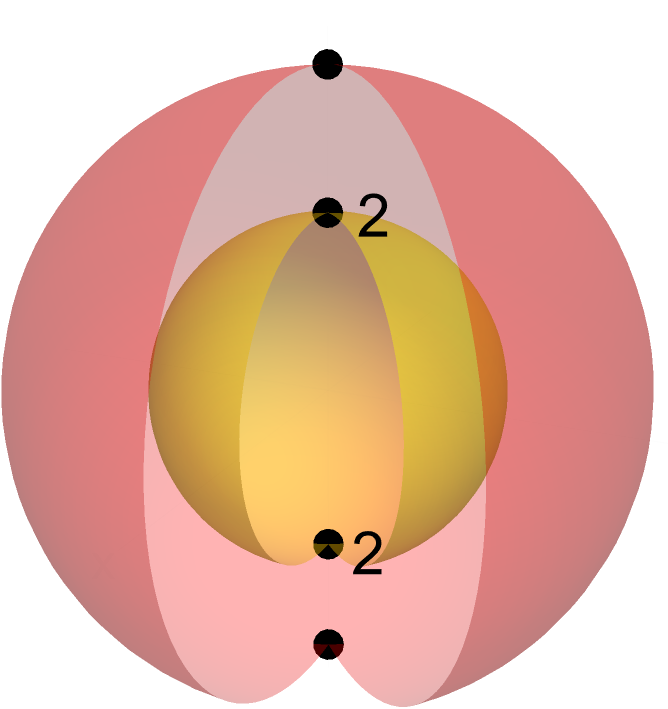}} 
\\  \hline
\scalebox{0.25}{\includegraphics{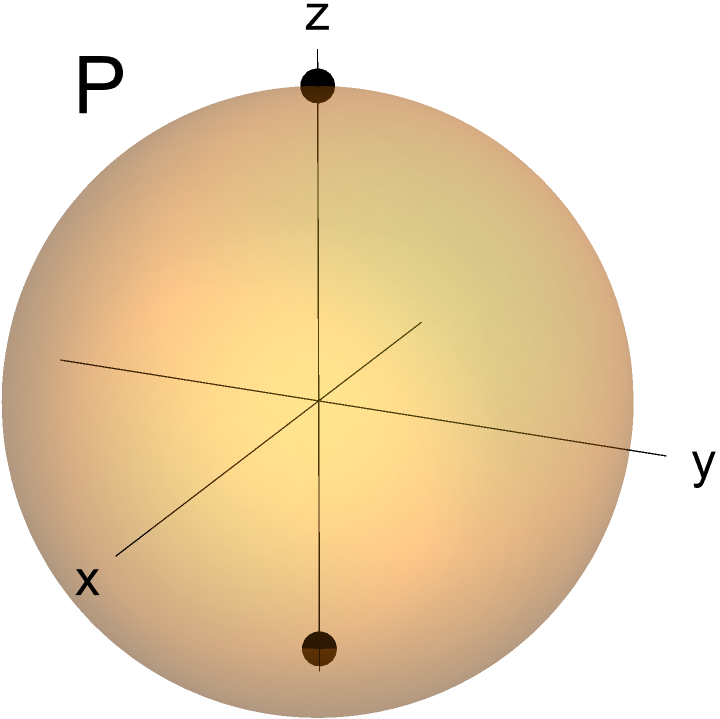}} & \scalebox{0.22}{\includegraphics{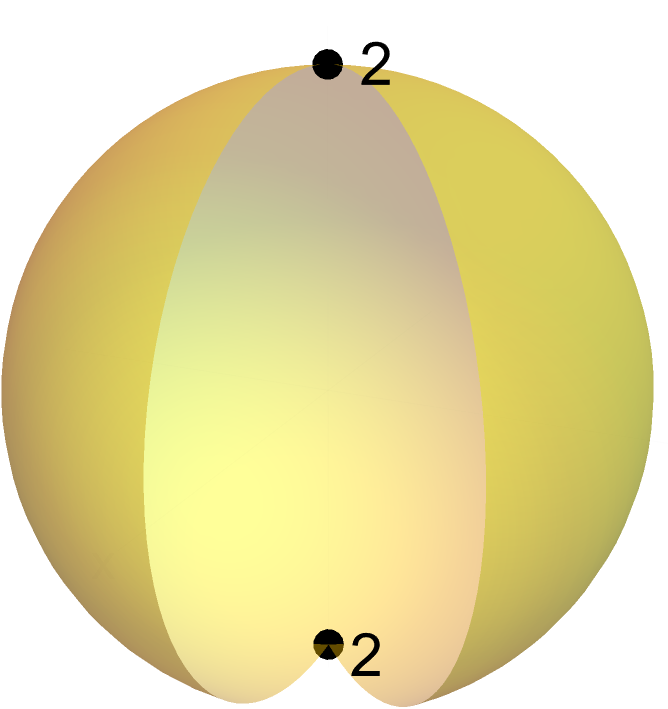}} 
\\  \hline 
\scalebox{0.2}{\includegraphics{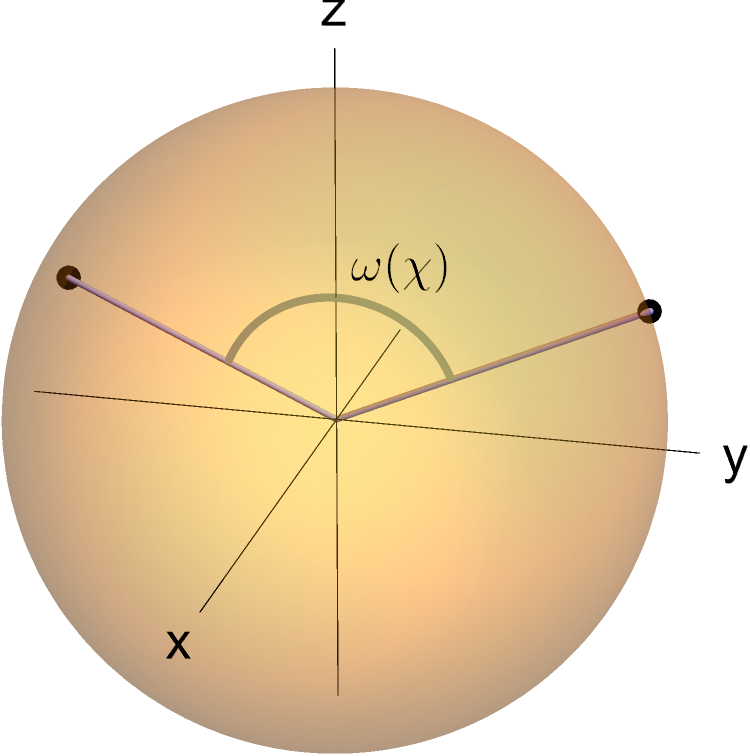}} & 
\scalebox{0.22}{\includegraphics{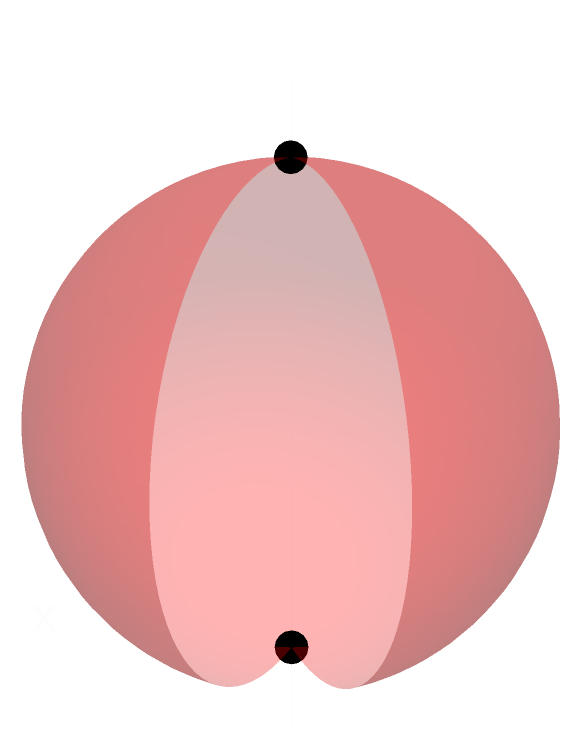}  \includegraphics{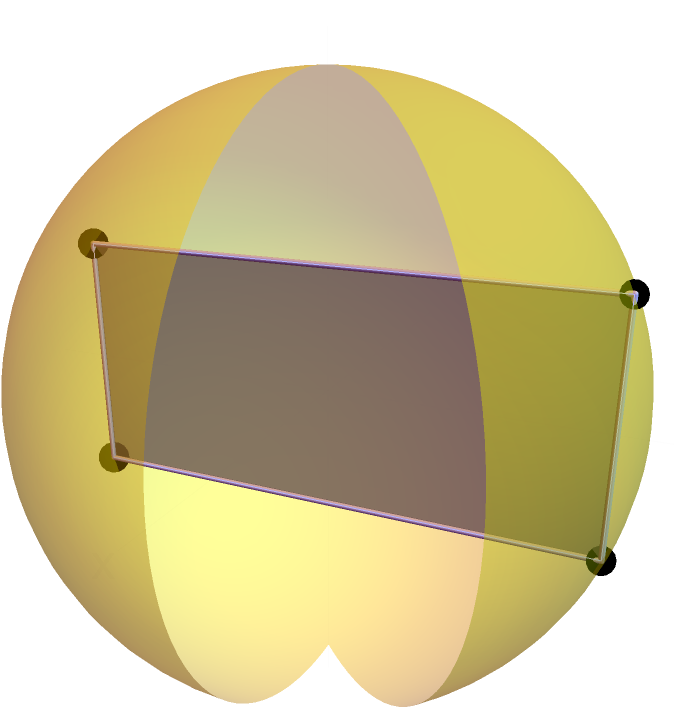}} 
\\  \hline
\end{tabular}
\caption{\label{MSC.f1} Majorana representations of the order parameters $\bm{\Phi}$ (left) and noncondensed fraction $\rho^{nc}$ (right) of spin phases of BEC of $f=1$. The adjacent number in some points correspond to its degeneracy. Each nonzero vector $\bm{\rho}_{\si}$ of $\rho^{nc}$ is associated to a constellation of $2\si$ points, which are also colored in red and yellow for $\si=1$ and $2$, respectively. For clarity, we omit the cartesian axis in the figures and split the constellations of the AF phase. The gray area in the second constellation of $\con{\rho^{nc}}$ of the AF phase is the corresponding geometric object of the points.
} 
\end{figure}

\begin{figure}[t!]
\large
\begin{tabular}{|c|c|}
\hline
 $\mathcal{C}_{\bm{\Phi}} $ & $\mathcal{C}_{\rho^{nc}}$
\\
\hline
\scalebox{0.22}{\includegraphics{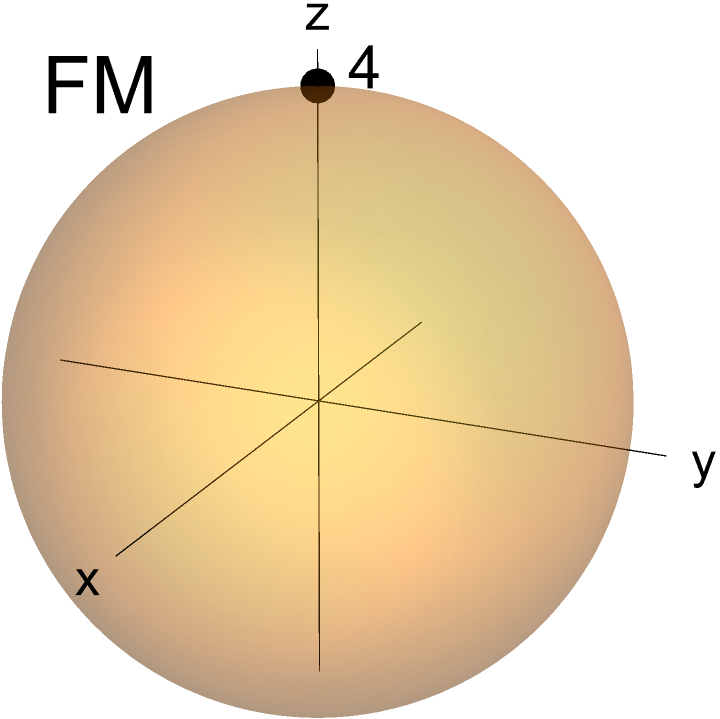}} 
& \scalebox{0.22}{\includegraphics{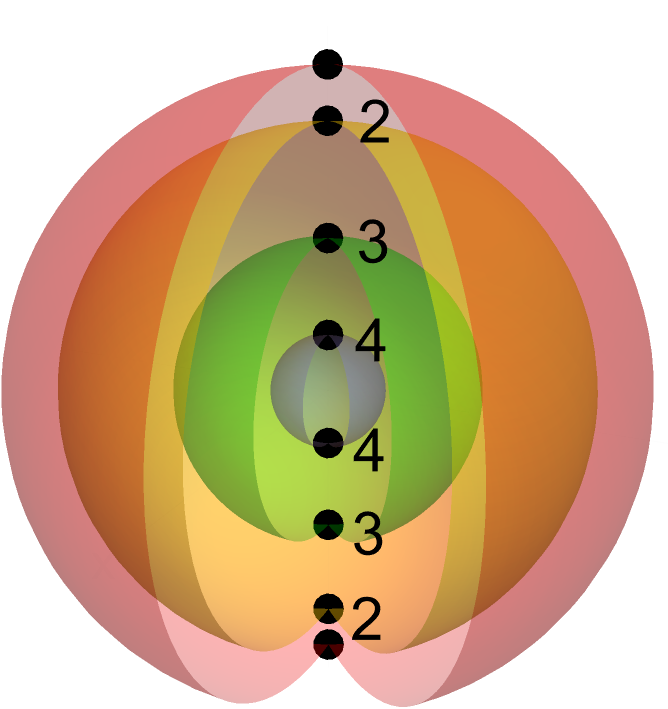}} 
\\
\hline
\scalebox{0.22}{\includegraphics{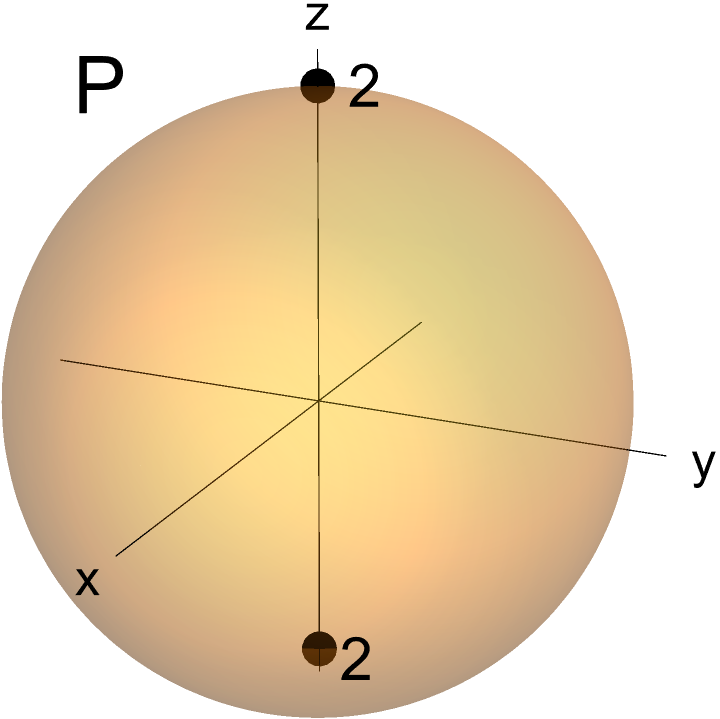}} & \scalebox{0.22}{\includegraphics{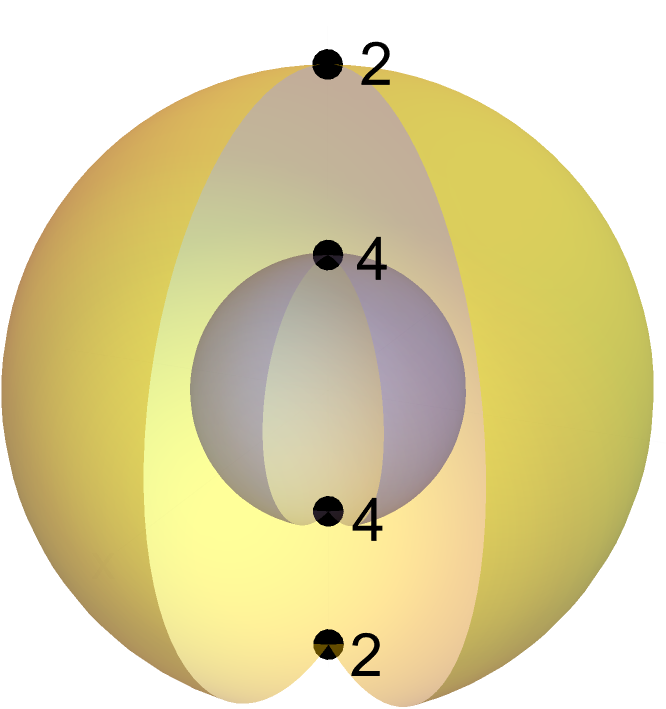}} 
\\
\hline
\scalebox{0.22}{\includegraphics{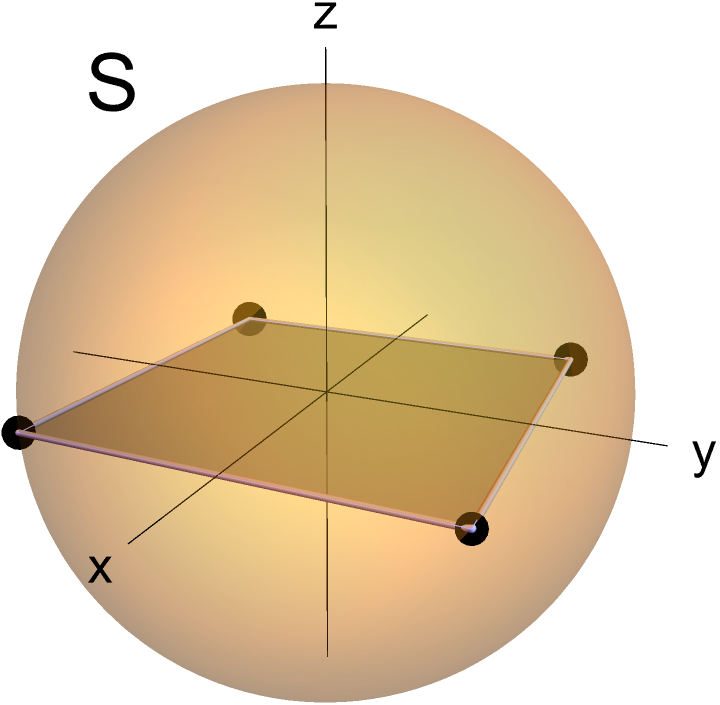}} &  \scalebox{0.22}{\includegraphics{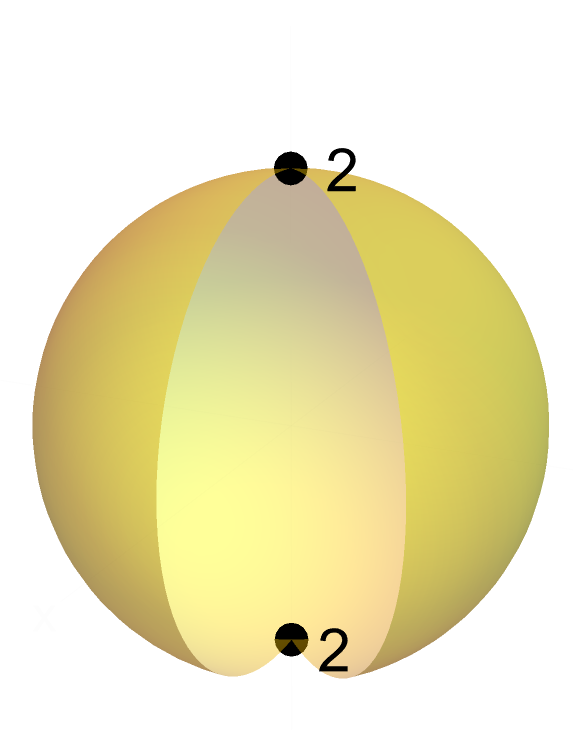} \includegraphics{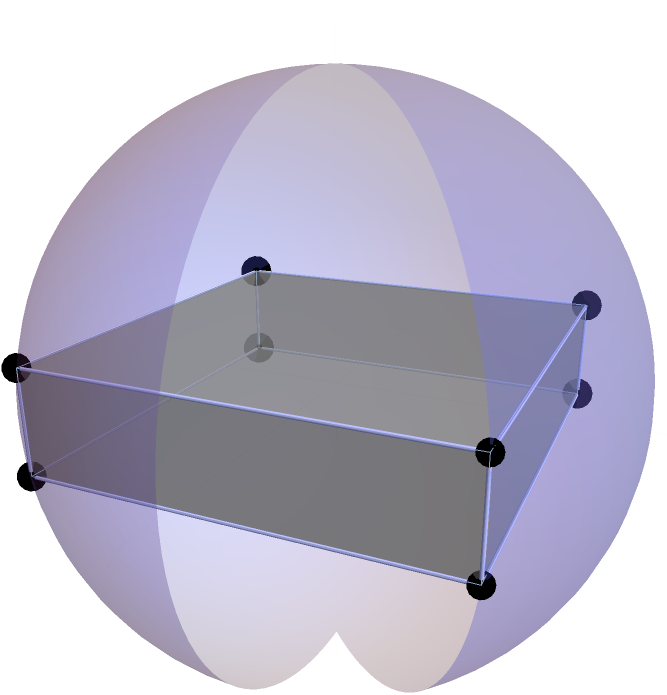}} 
\\
\hline
\scalebox{0.22}{\includegraphics{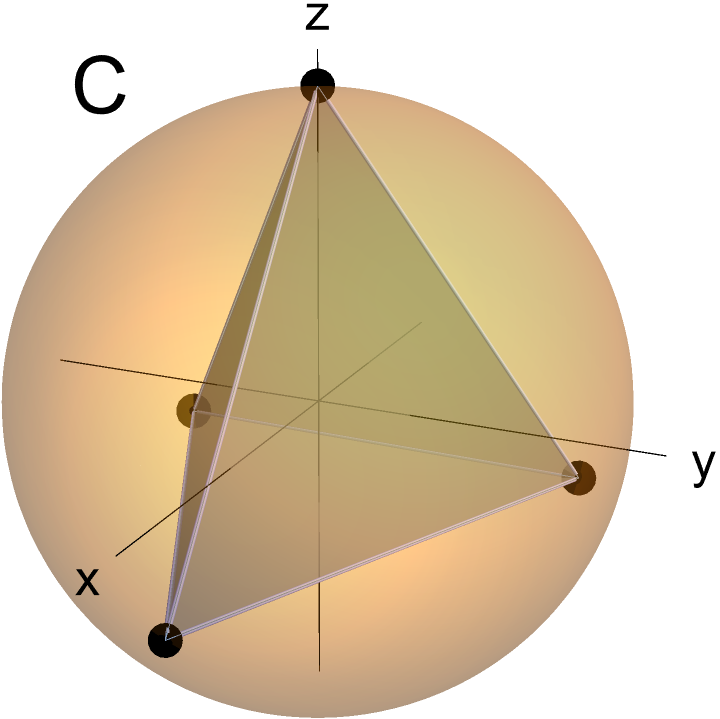}} & \scalebox{0.22}{\includegraphics{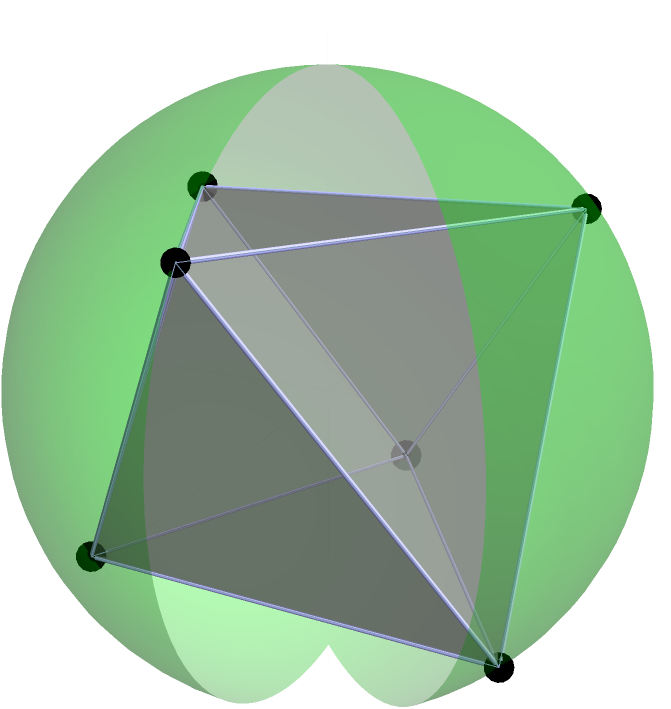} \includegraphics{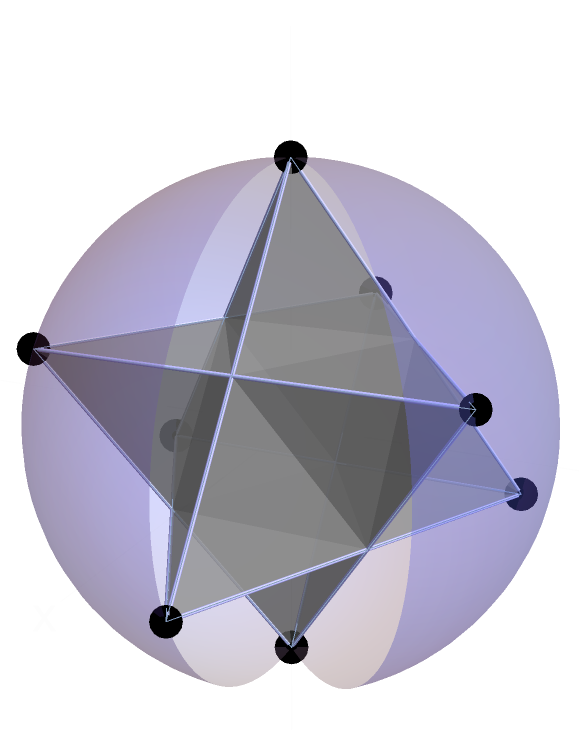}} 
\\
\hline
\end{tabular}
\caption{\label{MSC.f2} Majorana representations of $\bm{\Phi}$ (left) and $\rho^{nc}$ (right) of spin phases of BEC of spin-2, where we follow the same conventions as in Fig.~\ref{MSC.f1}. The constellations of $\bm{\rho}_{\si}$ for $\si=3,4$, correspond to the constellations with six (green sphere) and eight (blue sphere) points, respectively.
} 
\end{figure}

We now briefly review some of the most well-known spinor phases associated that corresponds to a MF solution of a BEC with spin $f=1,2$, and with a point group. In particular, we specify its order parameters and its corresponding Majorana constellations (Figs.~\ref{MSC.f1} and  \ref{MSC.f2}) for a given orientation:

\begin{enumerate}[label=(\roman*)]
\item Ferromagnetic (FM) phase: The spinor order-parameter has only one non-zero coefficient, $\phi_f = \sqrt{N}$. It is symmetric under rotations about the $z$ axis, imposing its symmetry group to be isomorphic to $SO(2)$.   Its constellation $\con{\bm{\Phi}}$ consists of $2f$ coincident points on the north pole.

\item Polar (P) phase: Here $\phi_m = \sqrt{N} \delta_{m0}$. Its symmetry group, $C_{\infty}$ in the Schönflies notation~\cite{book.bra.cra:10}, consists of the group generated by any rotation about the $z$ axis, and a rotation by $\pi$ about any axis on the equator. For spin $f=2$ condensates, it belongs to the family of states called the nematic phase~\cite{Bar.Tur.Dem:06}. The constellation of the polar phase has $f$ points on each pole of the sphere.

\item Antiferromagnetic (AF) phase: It is a non-inert state~\cite{Mak.Suo:07} since it is represented by a family of spin-1 states $\bm{\Phi} = \sqrt{N} ( \cos \chi , 0, \sin \chi)^{\text{T}}$ with $\chi \in (0,\pi/4]$. The whole family is symmetric over two geometric operations, a rotation by $\pi$ about the $z$ axis, and a reflection across the $yz$ plane, implying that the symmetry group is isomorphic to $\mathds{Z}_2 \times \mathds{Z}_2$. Its Majorana constellation consists of two points on the $yz$ plane, with an angle $\omega$ between them bisected by the $z$ axis (see Fig.~\ref{MSC.f1}). $\omega$ depends functionally on the parameter $\chi$ and increases monotonically. 

\item Square (S) phase: A spin-2 phase with non-zero order-parameter terms  $\phi_{2}= \phi_{-2} = \sqrt{N/2}$. Its Majorana constellation consists of a square. Hence, $\bm{\Phi}$ has the dihedral point group denoted by  $D_4$ in the Schönflies notation~\cite{book.bra.cra:10}. This phase belongs also to the family of the nematic spin-2 states~\cite{Bar.Tur.Dem:06}.

\item Cyclic (C) phase: This spin-2 phase is described with $\bm{\Phi}=(\sqrt{N/3})(1,0,0,\sqrt{2},0)^{\text T}$. The order parameter has a constellation equal to a tetrahedron with point group \emph{T} in the Schönflies notation \cite{book.bra.cra:10}.
\end{enumerate}

The point group of each spin state can be established by inspecting at its corresponding Majorana constellation. The spin phases mentioned above are still present as ground states for Hamiltonians with some additional symmetry breaking terms such as linear and quadratic Zeeman interactions, and fixed magnetization~\cite{Kaw.Ued:12,ciobanu2000phase}.

We now consider a variational perturbation $ \hat{\delta}_j$ of the field operators near the MF solution, such that $\hat{\psi}_j = \phi_j + \hat{\delta}_j$. The ultra-cold atomic gas is now described by two fractions: those who define the condensate ({\it c}) part, and those that specify the  noncondensate ({\it nc})  fraction of atoms. They are represented by the density matrices $\rho_{ij}^c = \phi_i \phi_j^* $ and $\rho_{ij}^{nc} = \langle \hat{\delta}_j^{\dagger} \hat{\delta}_i \rangle$, respectively. Here $N^a = \Tr (\rho^a)$ with $a=n , \, nc$ for the atomic fractions in each splitted part satisfying $N^c + N^{nc} = N$.
The ground state phases 
of the condensate will thus correspond to the spin states that minimizes the energy functional $E(\rho) = \langle \hat{H} \rangle$. By working within the Popov approximation, which assumes that all field operators of the form $\langle \hat{\delta}_j \hat{\delta}_i \rangle$ are ruled out, and further neglecting the contribution of the three-field operators of the form $\langle \hat{\delta}_k^{\dagger} \hat{\delta}_j
\hat{\delta}_i \rangle$~\cite{griffin2009bose}, we obtain for the energy,    
\begin{align}
& E(\rho) = \frac{c_0}{2} 
\left\{ 
N^2 + \Tr \left[ \rho^{nc} \left( 2\rho^c + \rho^{nc} \right) \right]
\right\}
\nonumber
\\
& + \frac{c_1}{2} \sum_{\alpha} \left\{
\Tr \left[ \rho F_{\alpha} \right]^2 + \Tr[F_{\alpha} \rho^{nc} F_{\alpha} (2\rho^c+ \rho^{nc})]
\right\}
\nonumber
\\
& + \frac{c_2}{10} \left\{ \Tr \left[
\TRS \rho \TRS \rho + \TRS \rho^{nc} \TRS \left( 2\rho^c + \rho^{nc} \right)
\right] \right\} \, ,
\label{Ener.fun}
\end{align}
here the coupling factors $c_k$ have physical units of energy per density square, and for spin-1 BEC, $c_2=0$. The time-reversal operator $\TRS$ acts in $\rho$ as~\cite{Ser.Bra:20}
\begin{equation}
(\TRS\rho \TRS)_{ji} = (-1)^{2f-i-j} \rho_{-i-j} \, .
\end{equation}
The previous equation leads to the MF energy~\cite{pethick2008bose,Kaw.Ued:12} in the assumption that $\rho_{ij}^{nc}=0$, and consequently $\rho= \rho^c $. In this case, the MF energy reduces to
\begin{equation}
E^{(MF)}(\rho)= \frac{c_0 N^2}{2} + \frac{c_1}{2} \sum_{\alpha} \Tr \left[ \rho F_{\alpha} \right]^2  + \frac{c_2}{10} \Tr \left[ \TRS \rho \TRS \rho \right]    \, ,
\end{equation}
In the scenario of a variational method with a {\it self-consistent symmetry}, $\rho^{nc}$ inherits the specific point group symmetries of the Hamiltonian~\eqref{Full.Ham} and $\rho^c$. We then need to determine the most general $\rho^{nc}$ for a given point group. To that end, we make use of the Majorana representation for mixed states~\cite{Ser.Bra:20}.
%
%
%

A mixed state is represented by a density matrix, \ie, a $(2f+1)\times (2f+1)$ complex matrix with a nonnegative eigenspectrum. The set of density matrices has an orthonormal basis given by the tensor operators $T^{(f)}_{\si \mu}$ with $\si=0,\dots 2f$ and $\mu = -\si , \dots , \si$ \cite{Fan:53,BrinkSatchler68,Var.Mos.Khe:88}, which are defined in terms of the Clebsch-Gordan
coefficients $C_{j_1 m_1 j_2 m_2}^{j m}$, and reads
\begin{equation}
\label{decomp.TensOp}
T_{\sigma \mu}^{(f)}
=
\sum_{m,m'=-f}^f (-1)^{f-m'} C_{fm,f-m'}^{\si \mu}\ket{f,m}\bra{f,m'} 
\, .
\end{equation}
For clarity, henceforth, we omit the super index $(f)$ when there is no possible confusion. The tensor operators $T_{\sigma \mu}$ satisfy
\begin{equation}
\Tr ( T_{\sigma_1 \mu_1}^{\dagger} T_{\sigma_2 \mu_2} ) = \delta_{\sigma_1 \sigma_2} \delta_{\mu_1 \mu_2} \, , \quad T_{\sigma \mu}^{\dagger} = (-1)^{\mu} T_{\sigma-\mu} \, .
\label{hermi.tens.op}
\end{equation} 
The most important property of the $T_{\si \mu }$ operators is that they transform  block-diagonally under a unitary transformation $U(\Rr)$, describing a rotation $\Rr \in SO(3)$ according to an \emph{irrep} of $SO(3)$,
 $D^{(\sigma)}(\Rr)$, that is
\begin{equation}
\label{prop.sh}
U(\Rr) T_{\sigma \mu} U^{-1}(\Rr) = \sum\limits_{\mu'=-\sigma}^{\sigma} D_{ \mu' \mu}^{(\sigma)}(\Rr) T_{\sigma \mu'} \, ,
\end{equation}
where $\sigma=0,1,2,\ldots$ labels the irrep. In the $T_{\sigma \mu}$ basis, the density matrix $\rho^{nc}$ can be written  as
\begin{equation}
\rho^{nc} = N^{nc} \left( \frac{\mathds{1}_f}{2f+1}+ \sum_{\si=1}^{2f} \bm{\rho}_{\si} \cdot \bm{T}_{\si} \right) \, ,
\label{bloc.decomp}
\end{equation}
where $\bm{\rho}_{\si}= (\rho_{\si \si} , \dots ,\rho_{\si -\si}) \in \mathds{C}^{2\si+1}$ with
$\rho_{\si \mu} = \Tr( \rho \, T^{\da}_{\si \mu})$, and $\bm{T}_{\si}=
(T_{\si \si} , \dots ,T_{\si , -\si})$ is a vector of matrices. Here the dot product stands for  $\sum_{\mu=-\sigma}^\sigma\rho_{\si \mu}T_{\si \mu} $. Each vector $\bm{\rho}_{\si}$, which transforms as a spinor of spin $\si$ by Eq.~\eqref{prop.sh}, can be associated to a constellation \emph{\`a la Majorana} \cite{Maj.Rep} consisting of $2\si$ points on $S^2$ obtained through a similar polynomial as Eq.~\eqref{first.pol} but defined with $\rho_{\si \mu}$. The hermiticity condition of $\rho^{nc}$ together with  Eq.~\eqref{hermi.tens.op} implies that every constellation $\co^{(\si)}$ has an antipodal symmetry \cite{Ser.Bra:20}. While a pure state $\bm{\Phi}$ is normalized and its global phase factor is physically irrelevant, the same quantities of $\bm{\rho}_{\si}$ carry now the necessary information to fully characterize $\rho^{nc}$. However, this information can also be added in the Majorana representation of $\rho^{nc}$. The norm of $\bm{\rho}_{\si}$, $r_{\si}$, is associated to the radius of the sphere where the constellation of $\bm{\rho}_{\si}$ lies. On the other hand, the hermiticity property of $\rho^{nc}$ implies that the global phase factor of $\bm{\rho}_{\si}$ can only have two choices \cite{Ser.Bra:20}, both differing by a minus sign. There exists a method to associate this sign to a certain equivalence class of the points of each constellation~\cite{Ser.Bra:20}. Here, we just incorporate this choice of sign to the norm $r_{\si}$. Hence, $r_{\si}$ can have negative values that evidently does not affect the radius of the sphere. In summary, a mixed state will be associated to a set of $2f$ constellations, denoted by $\con{\rho^{nc}}$,  with antipodal symmetry and a number of stars equal to $2\si$, with $\si=1,\dots , 2f$, over spheres with radii $r_{\si}$, respectively. 

We now determine the density matrices $\rho^{nc}$ with a particular point group $G$. By the property \eqref{prop.sh} of the Majorana representation, $\rho^{nc}$ has the point group $G$ if each $\bm{\rho}_{\si}$ fulfills
\begin{equation}
\label{Sym.cond}
D^{(\si)}(g) \bm{\rho}_{\si} = \bm{\rho}_{\si} \, , \quad \text{for each } \, g\in G \, .
\end{equation}
Let us remark that this condition is more restrictive that in the case of pure states, where a state $\bm{\Phi}$ is invariant under the element action $g\in G$ if $D(g) \bm{\Phi}$ is equal to $\bm{\Phi}$ up to a global phase factor. The determination of pure spin states with a particular point group has been studied before in \cite{Bag.Dam.Gir.Mar:15}. We use Eq.~\eqref{Sym.cond} to impose on $\rho^{nc}$ the symmetries of the spin phases mentioned above. We plot their Majorana representations in figures~\ref{MSC.f1} and \ref{MSC.f2}. By looking at the constellations, one can deduce that the point group of $\rho^{nc}$ is equal to their corresponding order parameter $\bm{\Phi}$.

\section{Results for spin-1 and 2 BECs}
\subsection{Characterization of the noncondensate fractions}
We use now the Majorana representation for mixed states to obtain the most general density matrix $\rho$ with a fixed point group Eq.~\eqref{Sym.cond}. In particular, we consider the point groups of the spin-1 and 2 phases mentioned above. Their respective Majorana constellations are shown in Figs.~\ref{MSC.f1} and \ref{MSC.f2}.
We summarize the most important characteristics of $\con{\rho^{nc}}$ of each phase:

\begin{enumerate}[label=(\roman*)]
\item[(i--ii)] FM and P phases:  $\bm{\rho}_{\si}$ has only the $0$th-components different than zero $\rho_{\si 0}= r_{\si}$. Their constellations are given by $\si$ stars on each pole of the sphere. The additional symmetry of the P phase implies that the $\bm{\rho}_{\si}$ vectors with $\si$ odd are zero.
\item[(iii)] AF phase: The vectors of $\bm{\rho}_{\si}$ are given by
\begin{align}
\bm{\rho}_1 = & \, r_1 \left( 0,1,0 \right) , \,
\nonumber
\\
\bm{\rho}_2 = & \, r_2 \left( \frac{\cos x}{\sqrt{2}} , 0 , \sin x , 0 , \frac{\cos x}{\sqrt{2}} \right) \, .
\end{align}
The constellations of $\bm{\rho}_1$ has a star on each pole, and for $\bm{\rho}_2$ it is a rectangle with sides parallels to the $y$ and $z$ axes with length dimensions dependent of the variable $x$.
\item[(iv)] S phase: $\rho^{nc}$ has only two non-zero vectors $\bm{\rho}_{\si}$
\begin{align}
  \bm{\rho}_2 = & \, r_2 \left(0 \, ,0 \, , 1 \, ,0 \, ,0 \right) \, ,
\nonumber
\\
\phantom{abab} \bm{\rho}_4 = & \, r_4 \left( \frac{\cos y}{\sqrt{2}} \, , 0 \, ,0 \, ,0 \, , \sin y \, ,0 \, ,0 \, ,0 \, , \frac{\cos y}{\sqrt{2}} \right) \, .
\end{align}
The constellation of $\bm{\rho}_2$ has two stars on each pole, and for $\bm{\rho}_4$ it consists of a parallelepiped with faces parallel to the cartesian planes and length dimensions dependent of the variable $y$.
\item[(v)] C phase: The $\bm{\rho}_{\si}$ non-zero vectors are 
\begin{align}
\bm{\rho}_3 = & \, r_3 \left(-\sqrt{2} \, ,0 \, ,0 \, , \sqrt{5} \, ,0 \, ,0 \, ,\sqrt{2} \right)/3 \, ,
\\
\phantom{abab} \bm{\rho}_4 = & \, r_4 \left( 0 \, , -\sqrt{10} \, ,0 \, ,0 \, , -\sqrt{7} \, ,0 \, ,0 \, , \sqrt{10} \, ,0 \right) / \sqrt{27}
\, .
\nonumber
\end{align} 
Their constellations are given by an octahedron and a constellation conformed by two antipodal tetrahedrons, respectively.
\end{enumerate}
In overall, a generic $\rho^{nc}$ will have a total of $(2f+1)^2$ degrees of freedom constituted by the variables $\rho_{\si \mu}$ and $N^{nc}$, with domain restricted by the properties of the density matrices $\rho^{nc}$, unit trace, hermiticity and positive semidefinite condition \cite{Bengtsson17}. Notwithstanding, the previous calculations show that the inherited symmetries of $\rho^{nc}$ reduce the degrees of freedom considerably. For example, in spin-$1$ BEC, the number is reduced from 9 degrees to 3 for both the FM and P phases, namely to $(N^{nc},r_1,r_2)$, and to 4 in the AF phase,  $(N^{nc},r_1,r_2,x)$. Additionally, the total degrees of freedom for the spin-2 phases, which add up to 25 in the general case, are drastically reduced to just 3 or 5, depending upon the symmetry of its corresponding order-parameter. 
\begin{figure}[t!]
\begin{center}
\scalebox{0.5}{
\includegraphics{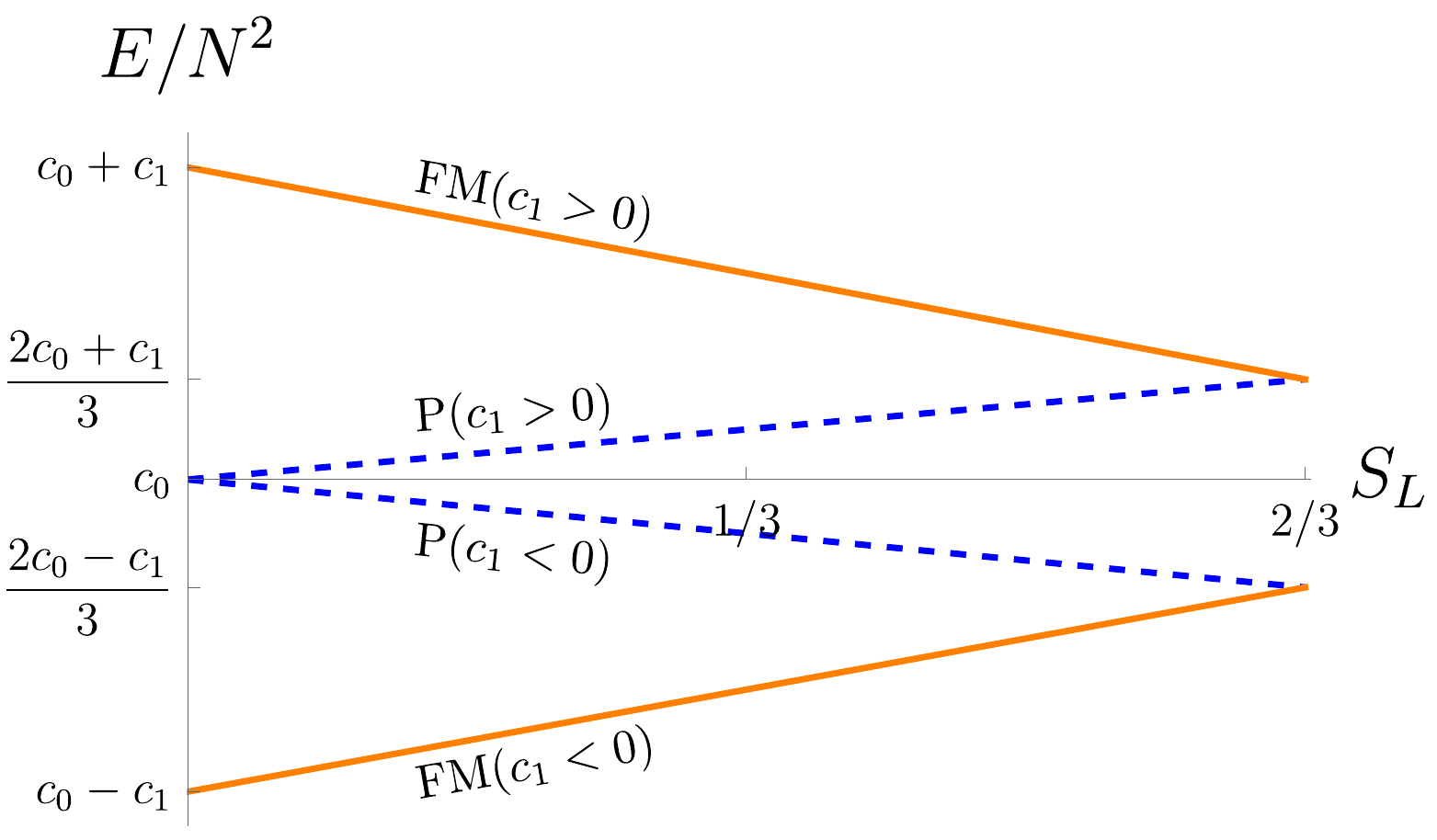}
} 

\scalebox{0.45}{
\hspace{0.6cm} \includegraphics{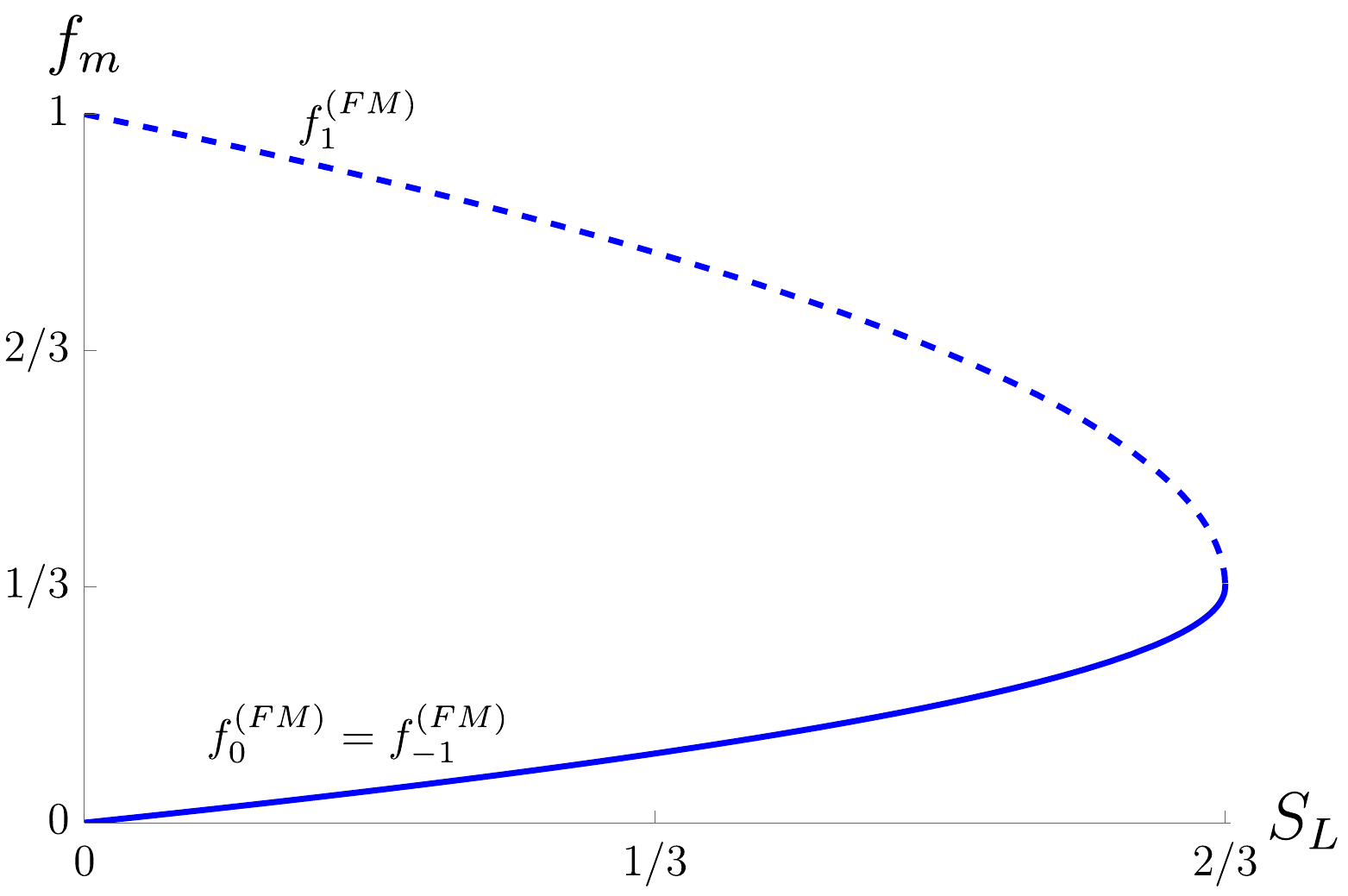}
}
\end{center}
\caption{\label{Fig.s1} 
(Top) Energy of the FM and P ground state phases $\rho$ with fixed linear entropy $S_L$ describing the mixture (disorder) of the system for positive and negative values of $c_1$. (Bottom) The fractions $f_m$ of the FM phase in the $\ket{s,m}$ states versus the linear entropy. For $S_L=0$, the system is in the MF solution of the FM phase, and as $S_L$ increases, the system tends  to the equally partitioned state achieved at $S_L \rightarrow 2/3$. Both limit cases illustrate the regimes $T=0$ and $T\rightarrow \infty$, respectively. The fractions of the P phase have a similar behavior (see Eq.~\eqref{frac.f1}).
}
\end{figure}
\subsection{Spin phases for $f=1$ and $f=2$}
For sake of illustration of our variational perturbation approach with {\it self-consistent symmetries}, we apply it to study the behavior of the phases for spin-1 and 2 BECs against temperature and entropy near the mean field solution. Let us first consider the spin-1 BEC case with spin phases at zero temperature having a nontrivial point group, namely, the FM, P, and AF phases, as described in Sec.~\ref{Sec.Met}. For the perturbation, instead to consider a particular scenario, let us just consider an increment of the entropy of the condensate, which is a signature of the atomic disorder (mixedness) of the condensate, due for instance to non-zero temperature effects. In that sense, we can study the physical properties of the spin-1 ground phases of a BEC with respect to the entropy given by~\cite{pathria2016statistical}
\begin{equation}
S(\rho)= -\Tr [(\rho/N) \ln (\rho/N) ] \, .
\end{equation}
Here, for simplicity, we work within the linear entropy response, obtained by Taylor expanding over the logarithm of a matrix around a pure state, yielding $S_L(\rho) \equiv  1 - \Tr \left[(\rho/N)^2 \right]$~\footnote{The logarithm of a mixed state ${\cal M}$ can be written in terms of its non-zero eigenvalues $m_k$ with eigenvector $\ket{ m_k}$, $\ln {\cal M}= \sum_k \ln m_k \ket{m_k}\bra{m_k}$. If the density matrix is almost a pure state, \ie~one of its eigenvalues $m_1$ is much larger than the others and closer to one, $\ln m_k $ is approximated by its first term of the Taylor expansion around the unit value $\ln m_k \approx m_k - 1$. Using this expansion and that $m_k \ll m_1$ for $k\neq 1$, we can obtain the approximation $-{\cal M}\ln {\cal M} \approx {\cal M}(\mathds{1}-{\cal M})$.}. It is noteworthy that $S_L(\rho)$ is a good approximation of the entropy $S(\rho)$ as long as 
$|S-S_L|\ll 1$. The previous regime can be written in terms of the non-zero eigenvalues $\La_k$ of $\rho$
\begin{equation}
 \left| 
1- \sum_{\La_k \neq 0} \frac{\La_k}{N} \left( \frac{\La_k}{N} - \ln \frac{\La_k}{N} \right)
 \right| \ll 1 \, .   
\end{equation}
Notice that for $T=0$, the density matrix has only one nonzero eigenvalue $\La_m=N$, yielding $S=S_L=0$. On the other hand, $T\rightarrow \infty$ entails, by the equipartition theorem, that the three eigenvalues are equal to $N/3$, and consequently $S_L = 2/3$. The ground spin phase is characterized by the state $\rho$ that minimizes the thermodynamical potential $\Omega(\rho) = E-Nk_BTS_L$. However, since we are going to consider a fixed value of the entropy $S_L$, the ground state phase is also identified as the density matrix  with minimal energy $E(\rho)$. The problem is then reduced to minimize Eq.~\eqref{Ener.fun} under the constraints of fixed number of particles $N = \Tr(\rho)$ and linear entropy $S_L(\rho)$, for the admissible phases of the spin-1 BEC. A direct calculation shows that the ground states allowed are only for the FM and the P phases for $c_1 <0$ and $c_1 >0$, respectively. The energies of the ground state phases have an analytical form, and are given by
\begin{align}
\nonumber
 E(\rho^{(FM)})= & N^2 \left[ c_0 \left(1- \frac{S_L}{2} \right) +c_1 (1 - S_L) \right] \, ,
\\
 E(\rho^{(P)})= & N^2 \left[ c_0 \left(1- \frac{S_L}{2} \right) + c_1 \frac{S_L}{2} \right] \, .
 \label{Energy.s1}
\end{align}
We plot the minimized energy $E(\rho)$ \eqref{Energy.s1} of the FM and P phases for a given value of $S_L(\rho)$ in Fig.~\ref{Fig.s1}. We can observe that for $c_1 >0$ $(c_1 <0)$, the ground state phase of the BEC has its atoms predominantly in the FM (P) phase. 
Therefore, our approach predicts that by an increment of the entropy in the spin-1 BEC, the FM-P phase  transition at $c_1=0$ does not exhibit a deviation. This observation agrees with other calculations found in the literature~\cite{Kawa.Phuc.Blakie:2012,Ser.Mir:21}, where in contrast, extensive numerical  calculations were considered instead to reach to the same conclusion.

Since the density matrices are diagonal with respect to the eigenbasis of the operator $F_z$, the condensate can thus be understood as consisting of a  statistical mixture of atoms in the $\ket{1,m}$ states. Each state is populated by a fraction of atoms equal to $f_m= \bra{1,\, m} \rho \ket{1, \, m}/N$, with values
\begin{align}
\nonumber
f_1^{(FM)} = &  f_0^{(P)}=\frac{1}{3} \left(\sqrt{4-6 S_L}+1\right) \, ,
\\
f_0^{(FM)} = &  f_{-1}^{(FM)} = f_1^{(P)} = f_{-1}^{(P)} = 
 \frac{1}{6} \left(2-\sqrt{4-6 S_L}\right) \, .
\label{frac.f1}
\end{align}
We plot the behavior of the fractions for the FM phases~\eqref{frac.f1} in Fig.~\ref{Fig.s1}. We can observe that for $S_L = 2/3$ ($T\rightarrow \infty$), the atoms of the BEC are equally distributed of the FM and P phases are equal, as expected by the equipartition theorem.
The qualitative behavior of the fractions exposed in Fig.~\eqref{frac.f1} is consistent with the numerical calculations reported in Ref.~\cite{Ser.Mir:21} where it is used the Hartree-Fock approximation. 
\begin{figure}[t!]
\scalebox{0.65}{\includegraphics{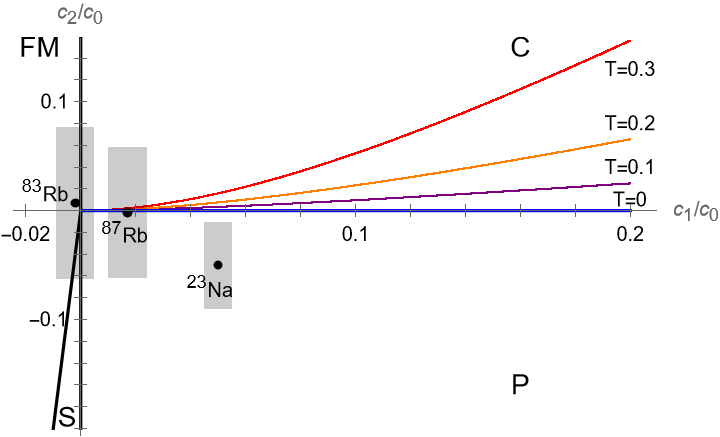}  } 
\caption{\label{Fig.4s2} 
Phase diagram of the spin-2 BEC in the space of the spin-dependent coupling factors $(c_1 , c_2)$. The C-P phase transition (color curves) depends on the temperature, while the rest of the phase transitions remain invariant (black lines). We also add the values of the coupling factors $(c_1 , c_2)$ of several atomic species along with its respective uncertainties \cite{ciobanu2000phase}.
} 
\end{figure}

The same procedure can be applied for a more sophisticated model. As an example, we apply the characterizations of $\rho^c$ and $\rho^{nc}$ to calculate the spin phase diagram of spin-2 BEC at finite temperatures using the Hartree-Fock approximation~\cite{blaizot1986quantum,griffin2009bose,Kawa.Phuc.Blakie:2012,Pro2008}. In this case, $\rho^{nc}$ is determined by the Bose-Einstein distribution at temperature $T$, while the spatial energy density is defined by its potential trap. For simplicity, we consider a finite box potential trap, with eigenstates labeled by the wavenumber vector $\bm{k}$, having a spatial energy contribution of $\hbar^2 k^2/2M$, being $M$ the atomic mass of the condensate. After the replacement $\sum_{\bm{k}}\rightarrow (2\pi)^{-3} \int\diff \bm{k}$, the spatial part of $\rho^{nc}$ is integrated leading to
\begin{equation}
\label{Poly.sta}
\rho^{nc}_{ij} = \sum_{\nu=1}^{2f+1} \xi^{\nu}_i \xi^{\nu *}_j \La_{\nu} \, , \quad \La_{\nu} =  \frac{Li_{3/2}\left(e^{-\beta \kappa_{\nu}} \right)}{\la_{dB}^3}  \, ,
\end{equation}
where $Li_{3/2}(z)$ is the polylogarithm function and $\la_{dB} = h / \sqrt{2\pi M k_B T} $ is the thermal de Broglie wavelength. The spin eigenstates $\xi^{\nu}$ with energies $\ka_{\nu}$ are defined by the eigendecomposition of the $(2f+1)\times (2f+1)$ matrix $A$, usually called the Hartree-Fock Hamiltonian, whose entries are derivatives of the energy~\eqref{Ener.fun}, $A_{ij} = \delta E / \delta \rho^{nc}_{ji}$~\cite{Ser.Mir:21}. In Fig.~\ref{Fig.4s2}, we plot the phase diagram of the spin-2 BECs at finite temperatures. We predict a deviation of the C-P phase-transition as a function of the temperature, in contrast with the other phase transitions that remain invariant with the temperature. 
We can conclude that, whilst for $^{23}$Na and $^{83}$Rb cold gases the phases remain practically insensible to temperature, the $^{87}$Rb condensates may exhibit a temperature-dependent spin-phase transition. Further details of the HF approximation applied to spin-1 and spin-2 BECs are discussed in \cite{Ser.Mir:21}. 
\section{Summary and Conclusions}
We have introduced a rigorous and systematic method capable of characterizing the MF variational perturbations of an interacting spinor system with a {\it self-consistent} rotational symmetry. Our method is based in a generalization of the Majorana stellar representation for quantum mixed states in conjunction with point group symmetry arguments. A distinctive feature of the approach is that it reduces considerably the unknown degrees of freedom of the perturbation. In addition the methodology can be applied to any other spin or spin-like physical system having {\it self-consistent} symmetries. We applied the approach to characterize the noncondensate fractions in a spin-1 BEC and to study the behavior of its spin phases when the mixedness of the condensate is monitored by the value of the entropy. The model is solved analytically and reproduces the same results predicted by others methods~\cite{Kawa.Phuc.Blakie:2012,Ser.Mir:21} but with much less intricacies involved. In addition, we calculate the phase diagram of spin-2 BEC in the Hartree-Fock approximation at finite temperatures, where a non-linear deviation in the C-P phase transition with temperature is predicted. The method presented here can be also be implemented in more complex systems, for example, for physical setups described by a tensor product of spins systems. However, further generalizations of the Majorana stellar representation may be necessary in such cases~\cite{PhysRevA.104.012407,chryssomalakos2021stellar}.
\\
\\

\section{Acknowledgments}
E.S.-E. acknowledges support from the postdoctoral fellowships offered by DGAPA-UNAM and the IPD-STEMA program of the University of Liège. F.M. acknowledges the support of DGAPA-UNAM through the project PAPIIT No. IN113920.
\bibliographystyle{apsrev4-2}
\bibliography{refs_joint_MajESE}

\begin{thebibliography}{63}%
\makeatletter
\providecommand \@ifxundefined [1]{%
 \@ifx{#1\undefined}
}%
\providecommand \@ifnum [1]{%
 \ifnum #1\expandafter \@firstoftwo
 \else \expandafter \@secondoftwo
 \fi
}%
\providecommand \@ifx [1]{%
 \ifx #1\expandafter \@firstoftwo
 \else \expandafter \@secondoftwo
 \fi
}%
\providecommand \natexlab [1]{#1}%
\providecommand \enquote  [1]{``#1''}%
\providecommand \bibnamefont  [1]{#1}%
\providecommand \bibfnamefont [1]{#1}%
\providecommand \citenamefont [1]{#1}%
\providecommand \href@noop [0]{\@secondoftwo}%
\providecommand \href [0]{\begingroup \@sanitize@url \@href}%
\providecommand \@href[1]{\@@startlink{#1}\@@href}%
\providecommand \@@href[1]{\endgroup#1\@@endlink}%
\providecommand \@sanitize@url [0]{\catcode `\\12\catcode `\$12\catcode
  `\&12\catcode `\#12\catcode `\^12\catcode `\_12\catcode `\%12\relax}%
\providecommand \@@startlink[1]{}%
\providecommand \@@endlink[0]{}%
\providecommand \url  [0]{\begingroup\@sanitize@url \@url }%
\providecommand \@url [1]{\endgroup\@href {#1}{\urlprefix }}%
\providecommand \urlprefix  [0]{URL }%
\providecommand \Eprint [0]{\href }%
\providecommand \doibase [0]{https://doi.org/}%
\providecommand \selectlanguage [0]{\@gobble}%
\providecommand \bibinfo  [0]{\@secondoftwo}%
\providecommand \bibfield  [0]{\@secondoftwo}%
\providecommand \translation [1]{[#1]}%
\providecommand \BibitemOpen [0]{}%
\providecommand \bibitemStop [0]{}%
\providecommand \bibitemNoStop [0]{.\EOS\space}%
\providecommand \EOS [0]{\spacefactor3000\relax}%
\providecommand \BibitemShut  [1]{\csname bibitem#1\endcsname}%
\let\auto@bib@innerbib\@empty
\bibitem [{\citenamefont {Lewenstein}\ \emph {et~al.}(2012)\citenamefont
  {Lewenstein}, \citenamefont {Sanpera},\ and\ \citenamefont
  {Ahufinger}}]{lewenstein2012ultracold}%
  \BibitemOpen
  \bibfield  {author} {\bibinfo {author} {\bibfnamefont {M.}~\bibnamefont
  {Lewenstein}}, \bibinfo {author} {\bibfnamefont {A.}~\bibnamefont
  {Sanpera}},\ and\ \bibinfo {author} {\bibfnamefont {V.}~\bibnamefont
  {Ahufinger}},\ }\href@noop {} {\emph {\bibinfo {title} {Ultracold Atoms in
  Optical Lattices: Simulating quantum many-body systems}}}\ (\bibinfo
  {publisher} {Oxford University Press},\ \bibinfo {year} {2012})\BibitemShut
  {NoStop}%
\bibitem [{\citenamefont {Kawaguchi}\ and\ \citenamefont
  {Ueda}(2012)}]{Kaw.Ued:12}%
  \BibitemOpen
  \bibfield  {author} {\bibinfo {author} {\bibfnamefont {Y.}~\bibnamefont
  {Kawaguchi}}\ and\ \bibinfo {author} {\bibfnamefont {M.}~\bibnamefont
  {Ueda}},\ }\href
  {https://doi.org/https://doi.org/10.1016/j.physrep.2012.07.005} {\bibfield
  {journal} {\bibinfo  {journal} {Phys. Rep.}\ }\textbf {\bibinfo {volume}
  {520}},\ \bibinfo {pages} {253 } (\bibinfo {year} {2012})}\BibitemShut
  {NoStop}%
\bibitem [{\citenamefont {Frapolli}\ \emph {et~al.}(2017)\citenamefont
  {Frapolli}, \citenamefont {Zibold}, \citenamefont {Invernizzi}, \citenamefont
  {Jim\'enez-Garc\'{\i}a}, \citenamefont {Dalibard},\ and\ \citenamefont
  {Gerbier}}]{PhysRevLett.119.050404}%
  \BibitemOpen
  \bibfield  {author} {\bibinfo {author} {\bibfnamefont {C.}~\bibnamefont
  {Frapolli}}, \bibinfo {author} {\bibfnamefont {T.}~\bibnamefont {Zibold}},
  \bibinfo {author} {\bibfnamefont {A.}~\bibnamefont {Invernizzi}}, \bibinfo
  {author} {\bibfnamefont {K.}~\bibnamefont {Jim\'enez-Garc\'{\i}a}}, \bibinfo
  {author} {\bibfnamefont {J.}~\bibnamefont {Dalibard}},\ and\ \bibinfo
  {author} {\bibfnamefont {F.}~\bibnamefont {Gerbier}},\ }\href
  {https://doi.org/10.1103/PhysRevLett.119.050404} {\bibfield  {journal}
  {\bibinfo  {journal} {Phys. Rev. Lett.}\ }\textbf {\bibinfo {volume} {119}},\
  \bibinfo {pages} {050404} (\bibinfo {year} {2017})}\BibitemShut {NoStop}%
\bibitem [{\citenamefont {Sch{\"a}fer}\ \emph {et~al.}(2020)\citenamefont
  {Sch{\"a}fer}, \citenamefont {Fukuhara}, \citenamefont {Sugawa},
  \citenamefont {Takasu},\ and\ \citenamefont {Takahashi}}]{schafer2020tools}%
  \BibitemOpen
  \bibfield  {author} {\bibinfo {author} {\bibfnamefont {F.}~\bibnamefont
  {Sch{\"a}fer}}, \bibinfo {author} {\bibfnamefont {T.}~\bibnamefont
  {Fukuhara}}, \bibinfo {author} {\bibfnamefont {S.}~\bibnamefont {Sugawa}},
  \bibinfo {author} {\bibfnamefont {Y.}~\bibnamefont {Takasu}},\ and\ \bibinfo
  {author} {\bibfnamefont {Y.}~\bibnamefont {Takahashi}},\ }\href@noop {}
  {\bibfield  {journal} {\bibinfo  {journal} {Nature Reviews Physics}\ }\textbf
  {\bibinfo {volume} {2}},\ \bibinfo {pages} {411} (\bibinfo {year}
  {2020})}\BibitemShut {NoStop}%
\bibitem [{\citenamefont {Tian}\ \emph {et~al.}(2020)\citenamefont {Tian},
  \citenamefont {Yang}, \citenamefont {Qiu}, \citenamefont {Liang},
  \citenamefont {Yang}, \citenamefont {Xu},\ and\ \citenamefont
  {Duan}}]{PRL.124.043001}%
  \BibitemOpen
  \bibfield  {author} {\bibinfo {author} {\bibfnamefont {T.}~\bibnamefont
  {Tian}}, \bibinfo {author} {\bibfnamefont {H.-X.}\ \bibnamefont {Yang}},
  \bibinfo {author} {\bibfnamefont {L.-Y.}\ \bibnamefont {Qiu}}, \bibinfo
  {author} {\bibfnamefont {H.-Y.}\ \bibnamefont {Liang}}, \bibinfo {author}
  {\bibfnamefont {Y.-B.}\ \bibnamefont {Yang}}, \bibinfo {author}
  {\bibfnamefont {Y.}~\bibnamefont {Xu}},\ and\ \bibinfo {author}
  {\bibfnamefont {L.-M.}\ \bibnamefont {Duan}},\ }\href
  {https://doi.org/10.1103/PhysRevLett.124.043001} {\bibfield  {journal}
  {\bibinfo  {journal} {Phys. Rev. Lett.}\ }\textbf {\bibinfo {volume} {124}},\
  \bibinfo {pages} {043001} (\bibinfo {year} {2020})}\BibitemShut {NoStop}%
\bibitem [{\citenamefont {Stenger}\ \emph {et~al.}(1998)\citenamefont
  {Stenger}, \citenamefont {Inouye}, \citenamefont {Stamper-Kurn},
  \citenamefont {Miesner}, \citenamefont {Chikkatur},\ and\ \citenamefont
  {Ketterle}}]{stenger1998spin}%
  \BibitemOpen
  \bibfield  {author} {\bibinfo {author} {\bibfnamefont {J.}~\bibnamefont
  {Stenger}}, \bibinfo {author} {\bibfnamefont {S.}~\bibnamefont {Inouye}},
  \bibinfo {author} {\bibfnamefont {D.}~\bibnamefont {Stamper-Kurn}}, \bibinfo
  {author} {\bibfnamefont {H.-J.}\ \bibnamefont {Miesner}}, \bibinfo {author}
  {\bibfnamefont {A.}~\bibnamefont {Chikkatur}},\ and\ \bibinfo {author}
  {\bibfnamefont {W.}~\bibnamefont {Ketterle}},\ }\href@noop {} {\bibfield
  {journal} {\bibinfo  {journal} {Nature}\ }\textbf {\bibinfo {volume} {396}},\
  \bibinfo {pages} {345} (\bibinfo {year} {1998})}\BibitemShut {NoStop}%
\bibitem [{\citenamefont {Stamper-Kurn}\ \emph {et~al.}(1998)\citenamefont
  {Stamper-Kurn}, \citenamefont {Andrews}, \citenamefont {Chikkatur},
  \citenamefont {Inouye}, \citenamefont {Miesner}, \citenamefont {Stenger},\
  and\ \citenamefont {Ketterle}}]{stamper.Andrews.etal:1998}%
  \BibitemOpen
  \bibfield  {author} {\bibinfo {author} {\bibfnamefont {D.~M.}\ \bibnamefont
  {Stamper-Kurn}}, \bibinfo {author} {\bibfnamefont {M.~R.}\ \bibnamefont
  {Andrews}}, \bibinfo {author} {\bibfnamefont {A.~P.}\ \bibnamefont
  {Chikkatur}}, \bibinfo {author} {\bibfnamefont {S.}~\bibnamefont {Inouye}},
  \bibinfo {author} {\bibfnamefont {H.-J.}\ \bibnamefont {Miesner}}, \bibinfo
  {author} {\bibfnamefont {J.}~\bibnamefont {Stenger}},\ and\ \bibinfo {author}
  {\bibfnamefont {W.}~\bibnamefont {Ketterle}},\ }\href@noop {} {\bibfield
  {journal} {\bibinfo  {journal} {Phys. Rev. Lett.}\ }\textbf {\bibinfo
  {volume} {80}},\ \bibinfo {pages} {2027} (\bibinfo {year}
  {1998})}\BibitemShut {NoStop}%
\bibitem [{\citenamefont {Jacob}\ \emph {et~al.}(2012)\citenamefont {Jacob},
  \citenamefont {Shao}, \citenamefont {Corre}, \citenamefont {Zibold},
  \citenamefont {De~Sarlo}, \citenamefont {Mimoun}, \citenamefont {Dalibard},\
  and\ \citenamefont {Gerbier}}]{Jacob:12}%
  \BibitemOpen
  \bibfield  {author} {\bibinfo {author} {\bibfnamefont {D.}~\bibnamefont
  {Jacob}}, \bibinfo {author} {\bibfnamefont {L.}~\bibnamefont {Shao}},
  \bibinfo {author} {\bibfnamefont {V.}~\bibnamefont {Corre}}, \bibinfo
  {author} {\bibfnamefont {T.}~\bibnamefont {Zibold}}, \bibinfo {author}
  {\bibfnamefont {L.}~\bibnamefont {De~Sarlo}}, \bibinfo {author}
  {\bibfnamefont {E.}~\bibnamefont {Mimoun}}, \bibinfo {author} {\bibfnamefont
  {J.}~\bibnamefont {Dalibard}},\ and\ \bibinfo {author} {\bibfnamefont
  {F.}~\bibnamefont {Gerbier}},\ }\href
  {https://doi.org/10.1103/PhysRevA.86.061601} {\bibfield  {journal} {\bibinfo
  {journal} {Phys. Rev. A}\ }\textbf {\bibinfo {volume} {86}},\ \bibinfo
  {pages} {061601(R)} (\bibinfo {year} {2012})}\BibitemShut {NoStop}%
\bibitem [{\citenamefont {Chang}\ \emph {et~al.}(2004)\citenamefont {Chang},
  \citenamefont {Hamley}, \citenamefont {Barrett}, \citenamefont {Sauer},
  \citenamefont {Fortier}, \citenamefont {Zhang}, \citenamefont {You},\ and\
  \citenamefont {Chapman}}]{Chang.Hamley.etal:2004}%
  \BibitemOpen
  \bibfield  {author} {\bibinfo {author} {\bibfnamefont {M.-S.}\ \bibnamefont
  {Chang}}, \bibinfo {author} {\bibfnamefont {C.~D.}\ \bibnamefont {Hamley}},
  \bibinfo {author} {\bibfnamefont {M.~D.}\ \bibnamefont {Barrett}}, \bibinfo
  {author} {\bibfnamefont {J.~A.}\ \bibnamefont {Sauer}}, \bibinfo {author}
  {\bibfnamefont {K.~M.}\ \bibnamefont {Fortier}}, \bibinfo {author}
  {\bibfnamefont {W.}~\bibnamefont {Zhang}}, \bibinfo {author} {\bibfnamefont
  {L.}~\bibnamefont {You}},\ and\ \bibinfo {author} {\bibfnamefont {M.~S.}\
  \bibnamefont {Chapman}},\ }\href
  {https://doi.org/10.1103/PhysRevLett.92.140403} {\bibfield  {journal}
  {\bibinfo  {journal} {Phys. Rev. Lett.}\ }\textbf {\bibinfo {volume} {92}},\
  \bibinfo {pages} {140403} (\bibinfo {year} {2004})}\BibitemShut {NoStop}%
\bibitem [{\citenamefont {Pethick}\ and\ \citenamefont
  {Smith}(2008)}]{pethick2008bose}%
  \BibitemOpen
  \bibfield  {author} {\bibinfo {author} {\bibfnamefont {C.~J.}\ \bibnamefont
  {Pethick}}\ and\ \bibinfo {author} {\bibfnamefont {H.}~\bibnamefont
  {Smith}},\ }\href@noop {} {\emph {\bibinfo {title} {Bose-{E}instein
  condensation in dilute gases}}}\ (\bibinfo  {publisher} {Cambridge University
  Press},\ \bibinfo {year} {2008})\BibitemShut {NoStop}%
\bibitem [{\citenamefont {Jim{\'e}nez-Garc{\'\i}a}\ \emph
  {et~al.}(2019)\citenamefont {Jim{\'e}nez-Garc{\'\i}a}, \citenamefont
  {Invernizzi}, \citenamefont {Evrard}, \citenamefont {Frapolli}, \citenamefont
  {Dalibard},\ and\ \citenamefont {Gerbier}}]{jimenez2019spontaneous}%
  \BibitemOpen
  \bibfield  {author} {\bibinfo {author} {\bibfnamefont {K.}~\bibnamefont
  {Jim{\'e}nez-Garc{\'\i}a}}, \bibinfo {author} {\bibfnamefont
  {A.}~\bibnamefont {Invernizzi}}, \bibinfo {author} {\bibfnamefont
  {B.}~\bibnamefont {Evrard}}, \bibinfo {author} {\bibfnamefont
  {C.}~\bibnamefont {Frapolli}}, \bibinfo {author} {\bibfnamefont
  {J.}~\bibnamefont {Dalibard}},\ and\ \bibinfo {author} {\bibfnamefont
  {F.}~\bibnamefont {Gerbier}},\ }\href@noop {} {\bibfield  {journal} {\bibinfo
   {journal} {Nat. Commun.}\ }\textbf {\bibinfo {volume} {10}},\ \bibinfo
  {pages} {1} (\bibinfo {year} {2019})}\BibitemShut {NoStop}%
\bibitem [{\citenamefont {Schmaljohann}\ \emph {et~al.}(2004)\citenamefont
  {Schmaljohann}, \citenamefont {Erhard}, \citenamefont {Kronj\"ager},
  \citenamefont {Kottke}, \citenamefont {van Staa}, \citenamefont
  {Cacciapuoti}, \citenamefont {Arlt}, \citenamefont {Bongs},\ and\
  \citenamefont {Sengstock}}]{Schmal.Erhard.etal:2004}%
  \BibitemOpen
  \bibfield  {author} {\bibinfo {author} {\bibfnamefont {H.}~\bibnamefont
  {Schmaljohann}}, \bibinfo {author} {\bibfnamefont {M.}~\bibnamefont
  {Erhard}}, \bibinfo {author} {\bibfnamefont {J.}~\bibnamefont {Kronj\"ager}},
  \bibinfo {author} {\bibfnamefont {M.}~\bibnamefont {Kottke}}, \bibinfo
  {author} {\bibfnamefont {S.}~\bibnamefont {van Staa}}, \bibinfo {author}
  {\bibfnamefont {L.}~\bibnamefont {Cacciapuoti}}, \bibinfo {author}
  {\bibfnamefont {J.~J.}\ \bibnamefont {Arlt}}, \bibinfo {author}
  {\bibfnamefont {K.}~\bibnamefont {Bongs}},\ and\ \bibinfo {author}
  {\bibfnamefont {K.}~\bibnamefont {Sengstock}},\ }\href
  {https://doi.org/10.1103/PhysRevLett.92.040402} {\bibfield  {journal}
  {\bibinfo  {journal} {Phys. Rev. Lett.}\ }\textbf {\bibinfo {volume} {92}},\
  \bibinfo {pages} {040402} (\bibinfo {year} {2004})}\BibitemShut {NoStop}%
\bibitem [{\citenamefont {Lu}\ \emph {et~al.}(2010)\citenamefont {Lu},
  \citenamefont {Youn},\ and\ \citenamefont {Lev}}]{PhysRevLett.104.063001}%
  \BibitemOpen
  \bibfield  {author} {\bibinfo {author} {\bibfnamefont {M.}~\bibnamefont
  {Lu}}, \bibinfo {author} {\bibfnamefont {S.~H.}\ \bibnamefont {Youn}},\ and\
  \bibinfo {author} {\bibfnamefont {B.~L.}\ \bibnamefont {Lev}},\ }\href
  {https://doi.org/10.1103/PhysRevLett.104.063001} {\bibfield  {journal}
  {\bibinfo  {journal} {Phys. Rev. Lett.}\ }\textbf {\bibinfo {volume} {104}},\
  \bibinfo {pages} {063001} (\bibinfo {year} {2010})}\BibitemShut {NoStop}%
\bibitem [{\citenamefont {McClelland}\ and\ \citenamefont
  {Hanssen}(2006)}]{PhysRevLett.96.143005}%
  \BibitemOpen
  \bibfield  {author} {\bibinfo {author} {\bibfnamefont {J.~J.}\ \bibnamefont
  {McClelland}}\ and\ \bibinfo {author} {\bibfnamefont {J.~L.}\ \bibnamefont
  {Hanssen}},\ }\href {https://doi.org/10.1103/PhysRevLett.96.143005}
  {\bibfield  {journal} {\bibinfo  {journal} {Phys. Rev. Lett.}\ }\textbf
  {\bibinfo {volume} {96}},\ \bibinfo {pages} {143005} (\bibinfo {year}
  {2006})}\BibitemShut {NoStop}%
\bibitem [{\citenamefont {Schreck}\ and\ \citenamefont
  {Druten}(2021)}]{schreck2021laser}%
  \BibitemOpen
  \bibfield  {author} {\bibinfo {author} {\bibfnamefont {F.}~\bibnamefont
  {Schreck}}\ and\ \bibinfo {author} {\bibfnamefont {K.~v.}\ \bibnamefont
  {Druten}},\ }\href@noop {} {\bibfield  {journal} {\bibinfo  {journal} {Nature
  Physics}\ }\textbf {\bibinfo {volume} {17}},\ \bibinfo {pages} {1296}
  (\bibinfo {year} {2021})}\BibitemShut {NoStop}%
\bibitem [{\citenamefont {Beaufils}\ \emph {et~al.}(2008)\citenamefont
  {Beaufils}, \citenamefont {Chicireanu}, \citenamefont {Zanon}, \citenamefont
  {Laburthe-Tolra}, \citenamefont {Mar\'echal}, \citenamefont {Vernac},
  \citenamefont {Keller},\ and\ \citenamefont {Gorceix}}]{PhysRevA.77.061601}%
  \BibitemOpen
  \bibfield  {author} {\bibinfo {author} {\bibfnamefont {Q.}~\bibnamefont
  {Beaufils}}, \bibinfo {author} {\bibfnamefont {R.}~\bibnamefont
  {Chicireanu}}, \bibinfo {author} {\bibfnamefont {T.}~\bibnamefont {Zanon}},
  \bibinfo {author} {\bibfnamefont {B.}~\bibnamefont {Laburthe-Tolra}},
  \bibinfo {author} {\bibfnamefont {E.}~\bibnamefont {Mar\'echal}}, \bibinfo
  {author} {\bibfnamefont {L.}~\bibnamefont {Vernac}}, \bibinfo {author}
  {\bibfnamefont {J.-C.}\ \bibnamefont {Keller}},\ and\ \bibinfo {author}
  {\bibfnamefont {O.}~\bibnamefont {Gorceix}},\ }\href
  {https://doi.org/10.1103/PhysRevA.77.061601} {\bibfield  {journal} {\bibinfo
  {journal} {Phys. Rev. A}\ }\textbf {\bibinfo {volume} {77}},\ \bibinfo
  {pages} {061601} (\bibinfo {year} {2008})}\BibitemShut {NoStop}%
\bibitem [{\citenamefont {Ho}(1998)}]{PhysRevLett.81.742}%
  \BibitemOpen
  \bibfield  {author} {\bibinfo {author} {\bibfnamefont {T.-L.}\ \bibnamefont
  {Ho}},\ }\href {https://doi.org/10.1103/PhysRevLett.81.742} {\bibfield
  {journal} {\bibinfo  {journal} {Phys. Rev. Lett.}\ }\textbf {\bibinfo
  {volume} {81}},\ \bibinfo {pages} {742} (\bibinfo {year} {1998})}\BibitemShut
  {NoStop}%
\bibitem [{\citenamefont {Ohmi}\ and\ \citenamefont
  {Machida}(1998)}]{ohmi1998bose}%
  \BibitemOpen
  \bibfield  {author} {\bibinfo {author} {\bibfnamefont {T.}~\bibnamefont
  {Ohmi}}\ and\ \bibinfo {author} {\bibfnamefont {K.}~\bibnamefont {Machida}},\
  }\href@noop {} {\bibfield  {journal} {\bibinfo  {journal} {J. Phys. Soc.}\
  }\textbf {\bibinfo {volume} {67}},\ \bibinfo {pages} {1822} (\bibinfo {year}
  {1998})}\BibitemShut {NoStop}%
\bibitem [{\citenamefont {Ciobanu}\ \emph {et~al.}(2000)\citenamefont
  {Ciobanu}, \citenamefont {Yip},\ and\ \citenamefont {Ho}}]{ciobanu2000phase}%
  \BibitemOpen
  \bibfield  {author} {\bibinfo {author} {\bibfnamefont {C.~V.}\ \bibnamefont
  {Ciobanu}}, \bibinfo {author} {\bibfnamefont {S.-K.}\ \bibnamefont {Yip}},\
  and\ \bibinfo {author} {\bibfnamefont {T.-L.}\ \bibnamefont {Ho}},\
  }\href@noop {} {\bibfield  {journal} {\bibinfo  {journal} {Phys. Rev. A}\
  }\textbf {\bibinfo {volume} {61}},\ \bibinfo {pages} {033607} (\bibinfo
  {year} {2000})}\BibitemShut {NoStop}%
\bibitem [{\citenamefont {Barnett}\ \emph {et~al.}(2006)\citenamefont
  {Barnett}, \citenamefont {Turner},\ and\ \citenamefont
  {Demler}}]{Bar.Tur.Dem:06}%
  \BibitemOpen
  \bibfield  {author} {\bibinfo {author} {\bibfnamefont {R.}~\bibnamefont
  {Barnett}}, \bibinfo {author} {\bibfnamefont {A.}~\bibnamefont {Turner}},\
  and\ \bibinfo {author} {\bibfnamefont {E.}~\bibnamefont {Demler}},\
  }\href@noop {} {\bibfield  {journal} {\bibinfo  {journal} {Phys. Rev. Lett.}\
  }\textbf {\bibinfo {volume} {97}},\ \bibinfo {pages} {180412} (\bibinfo
  {year} {2006})}\BibitemShut {NoStop}%
\bibitem [{\citenamefont {Diener}\ and\ \citenamefont
  {Ho}(2006)}]{diener2006cr}%
  \BibitemOpen
  \bibfield  {author} {\bibinfo {author} {\bibfnamefont {R.~B.}\ \bibnamefont
  {Diener}}\ and\ \bibinfo {author} {\bibfnamefont {T.-L.}\ \bibnamefont
  {Ho}},\ }\href@noop {} {\bibfield  {journal} {\bibinfo  {journal} {Phys. Rev.
  Lett.}\ }\textbf {\bibinfo {volume} {96}},\ \bibinfo {pages} {190405}
  (\bibinfo {year} {2006})}\BibitemShut {NoStop}%
\bibitem [{\citenamefont {Kawaguchi}\ and\ \citenamefont
  {Ueda}(2011)}]{PhysRevA.84.053616}%
  \BibitemOpen
  \bibfield  {author} {\bibinfo {author} {\bibfnamefont {Y.}~\bibnamefont
  {Kawaguchi}}\ and\ \bibinfo {author} {\bibfnamefont {M.}~\bibnamefont
  {Ueda}},\ }\href {https://doi.org/10.1103/PhysRevA.84.053616} {\bibfield
  {journal} {\bibinfo  {journal} {Phys. Rev. A}\ }\textbf {\bibinfo {volume}
  {84}},\ \bibinfo {pages} {053616} (\bibinfo {year} {2011})}\BibitemShut
  {NoStop}%
\bibitem [{\citenamefont {Kawaguchi}\ \emph {et~al.}(2012)\citenamefont
  {Kawaguchi}, \citenamefont {Phuc},\ and\ \citenamefont
  {Blakie}}]{Kawa.Phuc.Blakie:2012}%
  \BibitemOpen
  \bibfield  {author} {\bibinfo {author} {\bibfnamefont {Y.}~\bibnamefont
  {Kawaguchi}}, \bibinfo {author} {\bibfnamefont {N.~T.}\ \bibnamefont
  {Phuc}},\ and\ \bibinfo {author} {\bibfnamefont {P.~B.}\ \bibnamefont
  {Blakie}},\ }\href {https://doi.org/10.1103/PhysRevA.85.053611} {\bibfield
  {journal} {\bibinfo  {journal} {Phys. Rev. A}\ }\textbf {\bibinfo {volume}
  {85}},\ \bibinfo {pages} {053611} (\bibinfo {year} {2012})}\BibitemShut
  {NoStop}%
\bibitem [{\citenamefont {Jiang}\ \emph {et~al.}(2014)\citenamefont {Jiang},
  \citenamefont {Zhao}, \citenamefont {Webb},\ and\ \citenamefont
  {Liu}}]{PRA.90.023610}%
  \BibitemOpen
  \bibfield  {author} {\bibinfo {author} {\bibfnamefont {J.}~\bibnamefont
  {Jiang}}, \bibinfo {author} {\bibfnamefont {L.}~\bibnamefont {Zhao}},
  \bibinfo {author} {\bibfnamefont {M.}~\bibnamefont {Webb}},\ and\ \bibinfo
  {author} {\bibfnamefont {Y.}~\bibnamefont {Liu}},\ }\href
  {https://doi.org/10.1103/PhysRevA.90.023610} {\bibfield  {journal} {\bibinfo
  {journal} {Phys. Rev. A}\ }\textbf {\bibinfo {volume} {90}},\ \bibinfo
  {pages} {023610} (\bibinfo {year} {2014})}\BibitemShut {NoStop}%
\bibitem [{\citenamefont {Kang}\ \emph {et~al.}(2017)\citenamefont {Kang},
  \citenamefont {Seo}, \citenamefont {Kim},\ and\ \citenamefont
  {Shin}}]{PRA.95.053638}%
  \BibitemOpen
  \bibfield  {author} {\bibinfo {author} {\bibfnamefont {S.}~\bibnamefont
  {Kang}}, \bibinfo {author} {\bibfnamefont {S.~W.}\ \bibnamefont {Seo}},
  \bibinfo {author} {\bibfnamefont {J.~H.}\ \bibnamefont {Kim}},\ and\ \bibinfo
  {author} {\bibfnamefont {Y.}~\bibnamefont {Shin}},\ }\href
  {https://doi.org/10.1103/PhysRevA.95.053638} {\bibfield  {journal} {\bibinfo
  {journal} {Phys. Rev. A}\ }\textbf {\bibinfo {volume} {95}},\ \bibinfo
  {pages} {053638} (\bibinfo {year} {2017})}\BibitemShut {NoStop}%
\bibitem [{\citenamefont {Roy}\ \emph {et~al.}(2022)\citenamefont {Roy},
  \citenamefont {Gautam} \emph {et~al.}}]{roy2022collective}%
  \BibitemOpen
  \bibfield  {author} {\bibinfo {author} {\bibfnamefont {A.}~\bibnamefont
  {Roy}}, \bibinfo {author} {\bibfnamefont {S.}~\bibnamefont {Gautam}}, \emph
  {et~al.},\ }\href@noop {} {\bibfield  {journal} {\bibinfo  {journal} {arXiv
  preprint arXiv:2204.06898}\ } (\bibinfo {year} {2022})}\BibitemShut {NoStop}%
\bibitem [{\citenamefont {Serrano-Ens\'astiga}\ and\ \citenamefont
  {Mireles}(2021)}]{Ser.Mir:21}%
  \BibitemOpen
  \bibfield  {author} {\bibinfo {author} {\bibfnamefont {E.}~\bibnamefont
  {Serrano-Ens\'astiga}}\ and\ \bibinfo {author} {\bibfnamefont
  {F.}~\bibnamefont {Mireles}},\ }\href
  {https://doi.org/10.1103/PhysRevA.104.063308} {\bibfield  {journal} {\bibinfo
   {journal} {Phys. Rev. A}\ }\textbf {\bibinfo {volume} {104}},\ \bibinfo
  {pages} {063308} (\bibinfo {year} {2021})}\BibitemShut {NoStop}%
\bibitem [{\citenamefont {Esry}\ \emph {et~al.}(1997)\citenamefont {Esry},
  \citenamefont {Greene}, \citenamefont {Burke},\ and\ \citenamefont
  {Bohn}}]{PhysRevLett.78.3594}%
  \BibitemOpen
  \bibfield  {author} {\bibinfo {author} {\bibfnamefont {B.~D.}\ \bibnamefont
  {Esry}}, \bibinfo {author} {\bibfnamefont {C.~H.}\ \bibnamefont {Greene}},
  \bibinfo {author} {\bibfnamefont {J.~P.}\ \bibnamefont {Burke}, \bibfnamefont
  {Jr.}},\ and\ \bibinfo {author} {\bibfnamefont {J.~L.}\ \bibnamefont
  {Bohn}},\ }\href {https://doi.org/10.1103/PhysRevLett.78.3594} {\bibfield
  {journal} {\bibinfo  {journal} {Phys. Rev. Lett.}\ }\textbf {\bibinfo
  {volume} {78}},\ \bibinfo {pages} {3594} (\bibinfo {year}
  {1997})}\BibitemShut {NoStop}%
\bibitem [{\citenamefont {Shi}\ \emph {et~al.}(2000)\citenamefont {Shi},
  \citenamefont {Zheng},\ and\ \citenamefont {Chui}}]{PhysRevA.61.063613}%
  \BibitemOpen
  \bibfield  {author} {\bibinfo {author} {\bibfnamefont {H.}~\bibnamefont
  {Shi}}, \bibinfo {author} {\bibfnamefont {W.-M.}\ \bibnamefont {Zheng}},\
  and\ \bibinfo {author} {\bibfnamefont {S.-T.}\ \bibnamefont {Chui}},\ }\href
  {https://doi.org/10.1103/PhysRevA.61.063613} {\bibfield  {journal} {\bibinfo
  {journal} {Phys. Rev. A}\ }\textbf {\bibinfo {volume} {61}},\ \bibinfo
  {pages} {063613} (\bibinfo {year} {2000})}\BibitemShut {NoStop}%
\bibitem [{\citenamefont {Mueller}\ \emph {et~al.}(2006)\citenamefont
  {Mueller}, \citenamefont {Ho}, \citenamefont {Ueda},\ and\ \citenamefont
  {Baym}}]{PhysRevA.74.033612}%
  \BibitemOpen
  \bibfield  {author} {\bibinfo {author} {\bibfnamefont {E.~J.}\ \bibnamefont
  {Mueller}}, \bibinfo {author} {\bibfnamefont {T.-L.}\ \bibnamefont {Ho}},
  \bibinfo {author} {\bibfnamefont {M.}~\bibnamefont {Ueda}},\ and\ \bibinfo
  {author} {\bibfnamefont {G.}~\bibnamefont {Baym}},\ }\href
  {https://doi.org/10.1103/PhysRevA.74.033612} {\bibfield  {journal} {\bibinfo
  {journal} {Phys. Rev. A}\ }\textbf {\bibinfo {volume} {74}},\ \bibinfo
  {pages} {033612} (\bibinfo {year} {2006})}\BibitemShut {NoStop}%
\bibitem [{\citenamefont {Matuszewski}\ \emph {et~al.}(2008)\citenamefont
  {Matuszewski}, \citenamefont {Alexander},\ and\ \citenamefont
  {Kivshar}}]{PhysRevA.78.023632}%
  \BibitemOpen
  \bibfield  {author} {\bibinfo {author} {\bibfnamefont {M.}~\bibnamefont
  {Matuszewski}}, \bibinfo {author} {\bibfnamefont {T.~J.}\ \bibnamefont
  {Alexander}},\ and\ \bibinfo {author} {\bibfnamefont {Y.~S.}\ \bibnamefont
  {Kivshar}},\ }\href {https://doi.org/10.1103/PhysRevA.78.023632} {\bibfield
  {journal} {\bibinfo  {journal} {Phys. Rev. A}\ }\textbf {\bibinfo {volume}
  {78}},\ \bibinfo {pages} {023632} (\bibinfo {year} {2008})}\BibitemShut
  {NoStop}%
\bibitem [{\citenamefont {Mur-Petit}\ \emph {et~al.}(2006)\citenamefont
  {Mur-Petit}, \citenamefont {Guilleumas}, \citenamefont {Polls}, \citenamefont
  {Sanpera}, \citenamefont {Lewenstein}, \citenamefont {Bongs},\ and\
  \citenamefont {Sengstock}}]{PhysRevA.73.013629}%
  \BibitemOpen
  \bibfield  {author} {\bibinfo {author} {\bibfnamefont {J.}~\bibnamefont
  {Mur-Petit}}, \bibinfo {author} {\bibfnamefont {M.}~\bibnamefont
  {Guilleumas}}, \bibinfo {author} {\bibfnamefont {A.}~\bibnamefont {Polls}},
  \bibinfo {author} {\bibfnamefont {A.}~\bibnamefont {Sanpera}}, \bibinfo
  {author} {\bibfnamefont {M.}~\bibnamefont {Lewenstein}}, \bibinfo {author}
  {\bibfnamefont {K.}~\bibnamefont {Bongs}},\ and\ \bibinfo {author}
  {\bibfnamefont {K.}~\bibnamefont {Sengstock}},\ }\href
  {https://doi.org/10.1103/PhysRevA.73.013629} {\bibfield  {journal} {\bibinfo
  {journal} {Phys. Rev. A}\ }\textbf {\bibinfo {volume} {73}},\ \bibinfo
  {pages} {013629} (\bibinfo {year} {2006})}\BibitemShut {NoStop}%
\bibitem [{\citenamefont {Phuc}\ \emph {et~al.}(2013)\citenamefont {Phuc},
  \citenamefont {Kawaguchi},\ and\ \citenamefont {Ueda}}]{PhysRevA.88.043629}%
  \BibitemOpen
  \bibfield  {author} {\bibinfo {author} {\bibfnamefont {N.~T.}\ \bibnamefont
  {Phuc}}, \bibinfo {author} {\bibfnamefont {Y.}~\bibnamefont {Kawaguchi}},\
  and\ \bibinfo {author} {\bibfnamefont {M.}~\bibnamefont {Ueda}},\ }\href
  {https://doi.org/10.1103/PhysRevA.88.043629} {\bibfield  {journal} {\bibinfo
  {journal} {Phys. Rev. A}\ }\textbf {\bibinfo {volume} {88}},\ \bibinfo
  {pages} {043629} (\bibinfo {year} {2013})}\BibitemShut {NoStop}%
\bibitem [{\citenamefont {Vinit}\ \emph {et~al.}(2013)\citenamefont {Vinit},
  \citenamefont {Bookjans}, \citenamefont {S\'a~de Melo},\ and\ \citenamefont
  {Raman}}]{PRL.110.165301}%
  \BibitemOpen
  \bibfield  {author} {\bibinfo {author} {\bibfnamefont {A.}~\bibnamefont
  {Vinit}}, \bibinfo {author} {\bibfnamefont {E.~M.}\ \bibnamefont {Bookjans}},
  \bibinfo {author} {\bibfnamefont {C.~A.~R.}\ \bibnamefont {S\'a~de Melo}},\
  and\ \bibinfo {author} {\bibfnamefont {C.}~\bibnamefont {Raman}},\ }\href
  {https://doi.org/10.1103/PhysRevLett.110.165301} {\bibfield  {journal}
  {\bibinfo  {journal} {Phys. Rev. Lett.}\ }\textbf {\bibinfo {volume} {110}},\
  \bibinfo {pages} {165301} (\bibinfo {year} {2013})}\BibitemShut {NoStop}%
\bibitem [{\citenamefont {Shitara}\ \emph {et~al.}(2017)\citenamefont
  {Shitara}, \citenamefont {Bir},\ and\ \citenamefont {Blakie}}]{NJP.095003}%
  \BibitemOpen
  \bibfield  {author} {\bibinfo {author} {\bibfnamefont {N.}~\bibnamefont
  {Shitara}}, \bibinfo {author} {\bibfnamefont {S.}~\bibnamefont {Bir}},\ and\
  \bibinfo {author} {\bibfnamefont {P.~B.}\ \bibnamefont {Blakie}},\ }\href
  {https://doi.org/10.1088/1367-2630/aa7e70} {\bibfield  {journal} {\bibinfo
  {journal} {New Journal of Physics}\ }\textbf {\bibinfo {volume} {19}},\
  \bibinfo {pages} {095003} (\bibinfo {year} {2017})}\BibitemShut {NoStop}%
\bibitem [{\citenamefont {Symes}\ \emph {et~al.}(2018)\citenamefont {Symes},
  \citenamefont {Baillie},\ and\ \citenamefont {Blakie}}]{PRA.98.063618}%
  \BibitemOpen
  \bibfield  {author} {\bibinfo {author} {\bibfnamefont {L.~M.}\ \bibnamefont
  {Symes}}, \bibinfo {author} {\bibfnamefont {D.}~\bibnamefont {Baillie}},\
  and\ \bibinfo {author} {\bibfnamefont {P.~B.}\ \bibnamefont {Blakie}},\
  }\href {https://doi.org/10.1103/PhysRevA.98.063618} {\bibfield  {journal}
  {\bibinfo  {journal} {Phys. Rev. A}\ }\textbf {\bibinfo {volume} {98}},\
  \bibinfo {pages} {063618} (\bibinfo {year} {2018})}\BibitemShut {NoStop}%
\bibitem [{\citenamefont {Kim}\ \emph {et~al.}(2019)\citenamefont {Kim},
  \citenamefont {Hong}, \citenamefont {Kang},\ and\ \citenamefont
  {Shin}}]{PhysRevA.99.023606}%
  \BibitemOpen
  \bibfield  {author} {\bibinfo {author} {\bibfnamefont {J.~H.}\ \bibnamefont
  {Kim}}, \bibinfo {author} {\bibfnamefont {D.}~\bibnamefont {Hong}}, \bibinfo
  {author} {\bibfnamefont {S.}~\bibnamefont {Kang}},\ and\ \bibinfo {author}
  {\bibfnamefont {Y.}~\bibnamefont {Shin}},\ }\href
  {https://doi.org/10.1103/PhysRevA.99.023606} {\bibfield  {journal} {\bibinfo
  {journal} {Phys. Rev. A}\ }\textbf {\bibinfo {volume} {99}},\ \bibinfo
  {pages} {023606} (\bibinfo {year} {2019})}\BibitemShut {NoStop}%
\bibitem [{\citenamefont {Yang}\ \emph {et~al.}(2019)\citenamefont {Yang},
  \citenamefont {Tian}, \citenamefont {Yang}, \citenamefont {Qiu},
  \citenamefont {Liang}, \citenamefont {Chu}, \citenamefont
  {Da\ifmmode~\breve{g}\else \u{g}\fi{}}, \citenamefont {Xu}, \citenamefont
  {Liu},\ and\ \citenamefont {Duan}}]{PRA.100.013622}%
  \BibitemOpen
  \bibfield  {author} {\bibinfo {author} {\bibfnamefont {H.-X.}\ \bibnamefont
  {Yang}}, \bibinfo {author} {\bibfnamefont {T.}~\bibnamefont {Tian}}, \bibinfo
  {author} {\bibfnamefont {Y.-B.}\ \bibnamefont {Yang}}, \bibinfo {author}
  {\bibfnamefont {L.-Y.}\ \bibnamefont {Qiu}}, \bibinfo {author} {\bibfnamefont
  {H.-Y.}\ \bibnamefont {Liang}}, \bibinfo {author} {\bibfnamefont {A.-J.}\
  \bibnamefont {Chu}}, \bibinfo {author} {\bibfnamefont {C.~B.}\ \bibnamefont
  {Da\ifmmode~\breve{g}\else \u{g}\fi{}}}, \bibinfo {author} {\bibfnamefont
  {Y.}~\bibnamefont {Xu}}, \bibinfo {author} {\bibfnamefont {Y.}~\bibnamefont
  {Liu}},\ and\ \bibinfo {author} {\bibfnamefont {L.-M.}\ \bibnamefont
  {Duan}},\ }\href {https://doi.org/10.1103/PhysRevA.100.013622} {\bibfield
  {journal} {\bibinfo  {journal} {Phys. Rev. A}\ }\textbf {\bibinfo {volume}
  {100}},\ \bibinfo {pages} {013622} (\bibinfo {year} {2019})}\BibitemShut
  {NoStop}%
\bibitem [{\citenamefont {Blaizot}\ and\ \citenamefont
  {Ripka}(1986)}]{blaizot1986quantum}%
  \BibitemOpen
  \bibfield  {author} {\bibinfo {author} {\bibfnamefont {J.-P.}\ \bibnamefont
  {Blaizot}}\ and\ \bibinfo {author} {\bibfnamefont {G.}~\bibnamefont
  {Ripka}},\ }\href@noop {} {\emph {\bibinfo {title} {Quantum theory of finite
  systems}}}\ (\bibinfo  {publisher} {MIT press Cambridge, MA},\ \bibinfo
  {year} {1986})\BibitemShut {NoStop}%
\bibitem [{\citenamefont {Griffin}\ \emph {et~al.}(2009)\citenamefont
  {Griffin}, \citenamefont {Nikuni},\ and\ \citenamefont
  {Zaremba}}]{griffin2009bose}%
  \BibitemOpen
  \bibfield  {author} {\bibinfo {author} {\bibfnamefont {A.}~\bibnamefont
  {Griffin}}, \bibinfo {author} {\bibfnamefont {T.}~\bibnamefont {Nikuni}},\
  and\ \bibinfo {author} {\bibfnamefont {E.}~\bibnamefont {Zaremba}},\
  }\href@noop {} {\emph {\bibinfo {title} {Bose-condensed gases at finite
  temperatures}}}\ (\bibinfo  {publisher} {Cambridge University Press},\
  \bibinfo {year} {2009})\BibitemShut {NoStop}%
\bibitem [{\citenamefont {Proukakis}\ and\ \citenamefont
  {Jackson}(2008)}]{Pro2008}%
  \BibitemOpen
  \bibfield  {author} {\bibinfo {author} {\bibfnamefont {N.~P.}\ \bibnamefont
  {Proukakis}}\ and\ \bibinfo {author} {\bibfnamefont {B.}~\bibnamefont
  {Jackson}},\ }\href@noop {} {\bibfield  {journal} {\bibinfo  {journal}
  {Journal of Physics B: Atomic, Molecular and Optical Physics}\ }\textbf
  {\bibinfo {volume} {41}},\ \bibinfo {pages} {203002} (\bibinfo {year}
  {2008})}\BibitemShut {NoStop}%
\bibitem [{\citenamefont {P\'erez-Garc\'{\i}a}\ \emph
  {et~al.}(1997)\citenamefont {P\'erez-Garc\'{\i}a}, \citenamefont {Michinel},
  \citenamefont {Cirac}, \citenamefont {Lewenstein},\ and\ \citenamefont
  {Zoller}}]{Gar.Mic.Cir.Lew:97}%
  \BibitemOpen
  \bibfield  {author} {\bibinfo {author} {\bibfnamefont {V.~M.}\ \bibnamefont
  {P\'erez-Garc\'{\i}a}}, \bibinfo {author} {\bibfnamefont {H.}~\bibnamefont
  {Michinel}}, \bibinfo {author} {\bibfnamefont {J.~I.}\ \bibnamefont {Cirac}},
  \bibinfo {author} {\bibfnamefont {M.}~\bibnamefont {Lewenstein}},\ and\
  \bibinfo {author} {\bibfnamefont {P.}~\bibnamefont {Zoller}},\ }\href
  {https://doi.org/10.1103/PhysRevA.56.1424} {\bibfield  {journal} {\bibinfo
  {journal} {Phys. Rev. A}\ }\textbf {\bibinfo {volume} {56}},\ \bibinfo
  {pages} {1424} (\bibinfo {year} {1997})}\BibitemShut {NoStop}%
\bibitem [{\citenamefont {Hu}\ and\ \citenamefont
  {Liu}(2020)}]{PhysRevA.102.043302}%
  \BibitemOpen
  \bibfield  {author} {\bibinfo {author} {\bibfnamefont {H.}~\bibnamefont
  {Hu}}\ and\ \bibinfo {author} {\bibfnamefont {X.-J.}\ \bibnamefont {Liu}},\
  }\href {https://doi.org/10.1103/PhysRevA.102.043302} {\bibfield  {journal}
  {\bibinfo  {journal} {Phys. Rev. A}\ }\textbf {\bibinfo {volume} {102}},\
  \bibinfo {pages} {043302} (\bibinfo {year} {2020})}\BibitemShut {NoStop}%
\bibitem [{\citenamefont {Kanjilal}\ and\ \citenamefont
  {Bhattacharyay}(2022)}]{kanjilal2022variational}%
  \BibitemOpen
  \bibfield  {author} {\bibinfo {author} {\bibfnamefont {P.~K.}\ \bibnamefont
  {Kanjilal}}\ and\ \bibinfo {author} {\bibfnamefont {A.}~\bibnamefont
  {Bhattacharyay}},\ }\href@noop {} {\bibfield  {journal} {\bibinfo  {journal}
  {The European Physical Journal Plus}\ }\textbf {\bibinfo {volume} {137}},\
  \bibinfo {pages} {1} (\bibinfo {year} {2022})}\BibitemShut {NoStop}%
\bibitem [{\citenamefont {Michel}(1980)}]{RevModPhys.52.617}%
  \BibitemOpen
  \bibfield  {author} {\bibinfo {author} {\bibfnamefont {L.}~\bibnamefont
  {Michel}},\ }\href {https://doi.org/10.1103/RevModPhys.52.617} {\bibfield
  {journal} {\bibinfo  {journal} {Rev. Mod. Phys.}\ }\textbf {\bibinfo {volume}
  {52}},\ \bibinfo {pages} {617} (\bibinfo {year} {1980})}\BibitemShut
  {NoStop}%
\bibitem [{\citenamefont {M\"akel\"a}\ and\ \citenamefont
  {Suominen}(2007)}]{Mak.Suo:07}%
  \BibitemOpen
  \bibfield  {author} {\bibinfo {author} {\bibfnamefont {H.}~\bibnamefont
  {M\"akel\"a}}\ and\ \bibinfo {author} {\bibfnamefont {K.-A.}\ \bibnamefont
  {Suominen}},\ }\href {https://doi.org/10.1103/PhysRevLett.99.190408}
  {\bibfield  {journal} {\bibinfo  {journal} {Phys. Rev. Lett.}\ }\textbf
  {\bibinfo {volume} {99}},\ \bibinfo {pages} {190408} (\bibinfo {year}
  {2007})}\BibitemShut {NoStop}%
\bibitem [{\citenamefont {Yip}(2007)}]{PhysRevA.75.023625}%
  \BibitemOpen
  \bibfield  {author} {\bibinfo {author} {\bibfnamefont {S.-K.}\ \bibnamefont
  {Yip}},\ }\href {https://doi.org/10.1103/PhysRevA.75.023625} {\bibfield
  {journal} {\bibinfo  {journal} {Phys. Rev. A}\ }\textbf {\bibinfo {volume}
  {75}},\ \bibinfo {pages} {023625} (\bibinfo {year} {2007})}\BibitemShut
  {NoStop}%
\bibitem [{\citenamefont {Leanhardt}\ \emph {et~al.}(2002)\citenamefont
  {Leanhardt}, \citenamefont {G\"orlitz}, \citenamefont {Chikkatur},
  \citenamefont {Kielpinski}, \citenamefont {Shin}, \citenamefont {Pritchard},\
  and\ \citenamefont {Ketterle}}]{PhysRevLett.89.190403}%
  \BibitemOpen
  \bibfield  {author} {\bibinfo {author} {\bibfnamefont {A.~E.}\ \bibnamefont
  {Leanhardt}}, \bibinfo {author} {\bibfnamefont {A.}~\bibnamefont
  {G\"orlitz}}, \bibinfo {author} {\bibfnamefont {A.~P.}\ \bibnamefont
  {Chikkatur}}, \bibinfo {author} {\bibfnamefont {D.}~\bibnamefont
  {Kielpinski}}, \bibinfo {author} {\bibfnamefont {Y.}~\bibnamefont {Shin}},
  \bibinfo {author} {\bibfnamefont {D.~E.}\ \bibnamefont {Pritchard}},\ and\
  \bibinfo {author} {\bibfnamefont {W.}~\bibnamefont {Ketterle}},\ }\href
  {https://doi.org/10.1103/PhysRevLett.89.190403} {\bibfield  {journal}
  {\bibinfo  {journal} {Phys. Rev. Lett.}\ }\textbf {\bibinfo {volume} {89}},\
  \bibinfo {pages} {190403} (\bibinfo {year} {2002})}\BibitemShut {NoStop}%
\bibitem [{\citenamefont {Kasamatsu}\ \emph {et~al.}(2005)\citenamefont
  {Kasamatsu}, \citenamefont {Tsubota},\ and\ \citenamefont
  {Ueda}}]{kasamatsu2005vortices}%
  \BibitemOpen
  \bibfield  {author} {\bibinfo {author} {\bibfnamefont {K.}~\bibnamefont
  {Kasamatsu}}, \bibinfo {author} {\bibfnamefont {M.}~\bibnamefont {Tsubota}},\
  and\ \bibinfo {author} {\bibfnamefont {M.}~\bibnamefont {Ueda}},\ }\href@noop
  {} {\bibfield  {journal} {\bibinfo  {journal} {International Journal of
  Modern Physics B}\ }\textbf {\bibinfo {volume} {19}},\ \bibinfo {pages}
  {1835} (\bibinfo {year} {2005})}\BibitemShut {NoStop}%
\bibitem [{\citenamefont {Kobayashi}\ \emph {et~al.}(2009)\citenamefont
  {Kobayashi}, \citenamefont {Kawaguchi}, \citenamefont {Nitta},\ and\
  \citenamefont {Ueda}}]{Koba.Kawa.Nitta.Ueda:2009}%
  \BibitemOpen
  \bibfield  {author} {\bibinfo {author} {\bibfnamefont {M.}~\bibnamefont
  {Kobayashi}}, \bibinfo {author} {\bibfnamefont {Y.}~\bibnamefont
  {Kawaguchi}}, \bibinfo {author} {\bibfnamefont {M.}~\bibnamefont {Nitta}},\
  and\ \bibinfo {author} {\bibfnamefont {M.}~\bibnamefont {Ueda}},\ }\href
  {https://doi.org/10.1103/PhysRevLett.103.115301} {\bibfield  {journal}
  {\bibinfo  {journal} {Phys. Rev. Lett.}\ }\textbf {\bibinfo {volume} {103}},\
  \bibinfo {pages} {115301} (\bibinfo {year} {2009})}\BibitemShut {NoStop}%
\bibitem [{\citenamefont {Majorana}(1932)}]{Maj.Rep}%
  \BibitemOpen
  \bibfield  {author} {\bibinfo {author} {\bibfnamefont {E.}~\bibnamefont
  {Majorana}},\ }\href@noop {} {\bibfield  {journal} {\bibinfo  {journal}
  {Nuovo Cimento}\ }\textbf {\bibinfo {volume} {9}},\ \bibinfo {pages} {43}
  (\bibinfo {year} {1932})}\BibitemShut {NoStop}%
\bibitem [{\citenamefont {Serrano-Ens\'astiga}\ and\ \citenamefont
  {Braun}(2020)}]{Ser.Bra:20}%
  \BibitemOpen
  \bibfield  {author} {\bibinfo {author} {\bibfnamefont {E.}~\bibnamefont
  {Serrano-Ens\'astiga}}\ and\ \bibinfo {author} {\bibfnamefont
  {D.}~\bibnamefont {Braun}},\ }\href
  {https://doi.org/10.1103/PhysRevA.101.022332} {\bibfield  {journal} {\bibinfo
   {journal} {Phys. Rev. A}\ }\textbf {\bibinfo {volume} {101}},\ \bibinfo
  {pages} {022332} (\bibinfo {year} {2020})}\BibitemShut {NoStop}%
\bibitem [{\citenamefont {Chryssomalakos}\ \emph {et~al.}(2018)\citenamefont
  {Chryssomalakos}, \citenamefont {Guzm\'{a}n-Gonz\'{a}lez},\ and\
  \citenamefont {Serrano-Ens\'{a}stiga}}]{Chr.Guz.Ser:18}%
  \BibitemOpen
  \bibfield  {author} {\bibinfo {author} {\bibfnamefont {C.}~\bibnamefont
  {Chryssomalakos}}, \bibinfo {author} {\bibfnamefont {E.}~\bibnamefont
  {Guzm\'{a}n-Gonz\'{a}lez}},\ and\ \bibinfo {author} {\bibfnamefont
  {E.}~\bibnamefont {Serrano-Ens\'{a}stiga}},\ }\href@noop {} {\bibfield
  {journal} {\bibinfo  {journal} {J.{} Phys.{} A.}\ }\textbf {\bibinfo {volume}
  {51}},\ \bibinfo {pages} {165202} (\bibinfo {year} {2018})}\BibitemShut
  {NoStop}%
\bibitem [{\citenamefont {Varshalovich}\ \emph {et~al.}(1988)\citenamefont
  {Varshalovich}, \citenamefont {Moskalev},\ and\ \citenamefont
  {Khersonskii}}]{Var.Mos.Khe:88}%
  \BibitemOpen
  \bibfield  {author} {\bibinfo {author} {\bibfnamefont {D.}~\bibnamefont
  {Varshalovich}}, \bibinfo {author} {\bibfnamefont {A.}~\bibnamefont
  {Moskalev}},\ and\ \bibinfo {author} {\bibfnamefont {V.}~\bibnamefont
  {Khersonskii}},\ }\href@noop {} {\emph {\bibinfo {title} {Quantum {T}heory of
  {A}ngular {M}omentum}}}\ (\bibinfo  {publisher} {World Scientific},\ \bibinfo
  {year} {1988})\BibitemShut {NoStop}%
\bibitem [{\citenamefont {Bradley}\ and\ \citenamefont
  {Cracknell}(2010)}]{book.bra.cra:10}%
  \BibitemOpen
  \bibfield  {author} {\bibinfo {author} {\bibfnamefont {C.}~\bibnamefont
  {Bradley}}\ and\ \bibinfo {author} {\bibfnamefont {A.}~\bibnamefont
  {Cracknell}},\ }\href@noop {} {\emph {\bibinfo {title} {The mathematical
  theory of symmetry in solids: representation theory for point groups and
  space groups}}}\ (\bibinfo  {publisher} {Oxford University Press},\ \bibinfo
  {year} {2010})\BibitemShut {NoStop}%
\bibitem [{\citenamefont {Fano}(1953)}]{Fan:53}%
  \BibitemOpen
  \bibfield  {author} {\bibinfo {author} {\bibfnamefont {U.}~\bibnamefont
  {Fano}},\ }\href@noop {} {\bibfield  {journal} {\bibinfo  {journal} {Phys.
  Rev.}\ }\textbf {\bibinfo {volume} {90}},\ \bibinfo {pages} {577} (\bibinfo
  {year} {1953})}\BibitemShut {NoStop}%
\bibitem [{\citenamefont {Brink}\ and\ \citenamefont
  {Satchler}(1968)}]{BrinkSatchler68}%
  \BibitemOpen
  \bibfield  {author} {\bibinfo {author} {\bibfnamefont {D.}~\bibnamefont
  {Brink}}\ and\ \bibinfo {author} {\bibfnamefont {G.}~\bibnamefont
  {Satchler}},\ }\href@noop {} {\emph {\bibinfo {title} {Theory of Angular
  Momentum}}}\ (\bibinfo  {publisher} {Clarendon Press},\ \bibinfo {address}
  {Oxford},\ \bibinfo {year} {1968})\BibitemShut {NoStop}%
\bibitem [{\citenamefont {Baguette}\ \emph {et~al.}(2015)\citenamefont
  {Baguette}, \citenamefont {Damanet}, \citenamefont {Giraud},\ and\
  \citenamefont {Martin}}]{Bag.Dam.Gir.Mar:15}%
  \BibitemOpen
  \bibfield  {author} {\bibinfo {author} {\bibfnamefont {D.}~\bibnamefont
  {Baguette}}, \bibinfo {author} {\bibfnamefont {F.}~\bibnamefont {Damanet}},
  \bibinfo {author} {\bibfnamefont {O.}~\bibnamefont {Giraud}},\ and\ \bibinfo
  {author} {\bibfnamefont {J.}~\bibnamefont {Martin}},\ }\href
  {https://doi.org/10.1103/PhysRevA.92.052333} {\bibfield  {journal} {\bibinfo
  {journal} {Phys. Rev. A}\ }\textbf {\bibinfo {volume} {92}},\ \bibinfo
  {pages} {052333} (\bibinfo {year} {2015})}\BibitemShut {NoStop}%
\bibitem [{\citenamefont {I.Bengtsson}\ and\ \citenamefont
  {K.\.{Z}yczkowski}(2017)}]{Bengtsson17}%
  \BibitemOpen
  \bibfield  {author} {\bibinfo {author} {\bibnamefont {I.Bengtsson}}\ and\
  \bibinfo {author} {\bibnamefont {K.\.{Z}yczkowski}},\ }\href@noop {} {\emph
  {\bibinfo {title} {Geometry of quantum states: an introduction to quantum
  entanglement}}}\ (\bibinfo  {publisher} {Cambride University Press},\
  \bibinfo {year} {2017})\ \bibinfo {note} {2nd. Edition}\BibitemShut {NoStop}%
\bibitem [{\citenamefont {Pathria}(2016)}]{pathria2016statistical}%
  \BibitemOpen
  \bibfield  {author} {\bibinfo {author} {\bibfnamefont {R.~K.}\ \bibnamefont
  {Pathria}},\ }\href@noop {} {\emph {\bibinfo {title} {Statistical
  mechanics}}}\ (\bibinfo  {publisher} {Elsevier},\ \bibinfo {year}
  {2016})\BibitemShut {NoStop}%
\bibitem [{Note1()}]{Note1}%
  \BibitemOpen
  \bibinfo {note} {The logarithm of a mixed state ${\protect \cal M}$ can be
  written in terms of its non-zero eigenvalues $m_k$ with eigenvector
  $\mathinner {|{ m_k}\rangle }$, $\ln {\protect \cal M}= \DOTSB \sum@
  \slimits@ _k \ln m_k \mathinner {|{m_k}\rangle }\mathinner {\langle {m_k}|}$.
  If the density matrix is almost a pure state, \hbox {\protect \em i.e.{}}~one
  of its eigenvalues $m_1$ is much larger than the others and closer to one,
  $\ln m_k $ is approximated by its first term of the Taylor expansion around
  the unit value $\ln m_k \approx m_k - 1$. Using this expansion and that $m_k
  \ll m_1$ for $k\protect \neq 1$, we can obtain the approximation $-{\protect
  \cal M}\ln {\protect \cal M} \approx {\protect \cal M}(\protect \mathds
  {1}-{\protect \cal M})$.}\BibitemShut {Stop}%
\bibitem [{\citenamefont {Chryssomalakos}\ \emph
  {et~al.}(2021{\natexlab{a}})\citenamefont {Chryssomalakos}, \citenamefont
  {Hanotel}, \citenamefont {Guzm\'an-Gonz\'alez}, \citenamefont {Braun},
  \citenamefont {Serrano-Ens\'astiga},\ and\ \citenamefont
  {\ifmmode~\dot{Z}\else \.{Z}\fi{}yczkowski}}]{PhysRevA.104.012407}%
  \BibitemOpen
  \bibfield  {author} {\bibinfo {author} {\bibfnamefont {C.}~\bibnamefont
  {Chryssomalakos}}, \bibinfo {author} {\bibfnamefont {L.}~\bibnamefont
  {Hanotel}}, \bibinfo {author} {\bibfnamefont {E.}~\bibnamefont
  {Guzm\'an-Gonz\'alez}}, \bibinfo {author} {\bibfnamefont {D.}~\bibnamefont
  {Braun}}, \bibinfo {author} {\bibfnamefont {E.}~\bibnamefont
  {Serrano-Ens\'astiga}},\ and\ \bibinfo {author} {\bibfnamefont
  {K.}~\bibnamefont {\ifmmode~\dot{Z}\else \.{Z}\fi{}yczkowski}},\ }\href
  {https://doi.org/10.1103/PhysRevA.104.012407} {\bibfield  {journal} {\bibinfo
   {journal} {Phys. Rev. A}\ }\textbf {\bibinfo {volume} {104}},\ \bibinfo
  {pages} {012407} (\bibinfo {year} {2021}{\natexlab{a}})}\BibitemShut
  {NoStop}%
\bibitem [{\citenamefont {Chryssomalakos}\ \emph
  {et~al.}(2021{\natexlab{b}})\citenamefont {Chryssomalakos}, \citenamefont
  {Guzm{\'a}n-Gonz{\'a}lez}, \citenamefont {Hanotel},\ and\ \citenamefont
  {Serrano-Ens{\'a}stiga}}]{chryssomalakos2021stellar}%
  \BibitemOpen
  \bibfield  {author} {\bibinfo {author} {\bibfnamefont {C.}~\bibnamefont
  {Chryssomalakos}}, \bibinfo {author} {\bibfnamefont {E.}~\bibnamefont
  {Guzm{\'a}n-Gonz{\'a}lez}}, \bibinfo {author} {\bibfnamefont
  {L.}~\bibnamefont {Hanotel}},\ and\ \bibinfo {author} {\bibfnamefont
  {E.}~\bibnamefont {Serrano-Ens{\'a}stiga}},\ }\href@noop {} {\bibfield
  {journal} {\bibinfo  {journal} {Communications in Mathematical Physics}\
  }\textbf {\bibinfo {volume} {381}},\ \bibinfo {pages} {735} (\bibinfo {year}
  {2021}{\natexlab{b}})}\BibitemShut {NoStop}%
\end{thebibliography}%
\end{document}